\RequirePackage{fix-cm}
\documentclass[smallextended]{svjour3}       
\smartqed  
\usepackage[sort&compress]{natbib}
\usepackage{graphicx}
\usepackage{floatrow}
\usepackage{siunitx}
\usepackage{amsmath}
\usepackage{comment}
\usepackage{amssymb}	
\usepackage{color}
\usepackage{soul}
\usepackage{arydshln}
\usepackage{array}
\newcolumntype{L}[1]{>{\raggedright\let\newline\\\arraybackslash\hspace{0pt}}p{#1}}
\newcolumntype{C}[1]{>{\centering\let\newline\\\arraybackslash\hspace{0pt}}p{#1}}
\newcolumntype{R}[1]{>{\raggedleft\let\newline\\\arraybackslash\hspace{0pt}}p{#1}}
\allowdisplaybreaks[4]
 \journalname{Space Science Reviews}
\begin{document}

\title{Anthropogenic Space Weather}

\titlerunning{Anthropogenic Space Weather}        

\author{T. I. Gombosi \and
        D. N. Baker \and
        A. Balogh \and
        P. J. Erickson \and
        J. D. Huba \and
        L. J. Lanzerotti 
}


\institute{Tamas Gombosi \at
               University of Michigan, Ann Arbor, MI 48109, USA;
               \email{tamas@umich.edu}           
           \and
           Daniel Baker \at
              University of Colorado, Boulder, CO 80303, USA
           \and
           Andr{\'e} Balogh \at
              Imperial College, London, United Kingdom
           \and
           Philip Erickson \at
              MIT Haystack Observatory, Westford, MA 01886, USA
           \and
           Joseph Huba \at
              Naval Research Laboratory, Washington, DC 20375, USA
           \and
           Louis Lanzerotti \at
              New Jersey Institute of Technology, Newark, NJ 07102, USA and \at
              Alcatel Lucent Bell Laboratories, Murray Hill, NJ 07974, USA (ret)
}

\date{Received: \today/ Accepted: \today}

\maketitle

%
%
\begin{abstract}
Anthropogenic effects on the space environment started in the late 19th century and reached their peak in the 1960s when high-altitude nuclear explosions were carried out by the USA and the Soviet Union. These explosions created artificial radiation belts near Earth that resulted in major damages to several satellites. Another, unexpected impact of the high-altitude nuclear tests was the electromagnetic pulse (EMP) that can have devastating effects over a large geographic area (as large as the continental United States). Other anthropogenic impacts on the space environment include chemical release experiments, high-frequency wave heating of the ionosphere and the interaction of VLF waves with the radiation belts. This paper reviews the fundamental physical process behind these phenomena and discusses the observations of their impacts.
\keywords{High-altitude nuclear explosions \and 
Artificial radiation belts \and 
Electromagnetic pulse (EMP) \and
Damage to satellites \and
Space Debris \and
Chemical releases \and
HF heating \and
VLF waves and radiation belts}
\PACS{94.05.S- \and 94.05.sj \and 94.05.sk \and 94.05.sy \and 94.20.wf \and 94.20.ws \and 94.30.Xy}
\end{abstract}


\section{Introduction}
\label{intro}

The ``victory'' of Nikola Tesla's alternating current (AC) as implemented by George Westinghouse (to whom Tesla sold most of his patents) over Thomas Edison's direct current (DC) as the means to power and to light up the U.S. and the world beginning in the late 19th century can be considered the initiation point of potential anthropogenic modifications of Earth's space environment.    However, it was not until nearly a century later when the very quiet electromagnetic environment of Siple Station, Antarctica, came online that the evidence of power line harmonic radiation was discovered in Earth's magnetosphere \cite[]{Helliwell:1975}.   

Since this report by the Stanford group, a considerable number of ground- and space-based studies have been published that have examined the potential effects of these harmonics in the VLF range on the electron population of the magnetosphere.  The published discussions have ranged from ``control'' of the magnetosphere through actions on, or production of, chorus emissions; \cite[e.g.][]{Park:1977, Luette:1979, Bullough:1983, Parrot:1994} to considerable skepticism \cite[e.g.][]{Tsurutani:1979, Tsurutani:1981}.   The possible effects on the space environment of human activities via use of electrical power sources in industrial activities have also been reported from statistical analyses of weekly variations in geomagnetic activity \cite[e.g.][]{Park:1979, Fraser-Smith:1979}, and refuted from other analyses \cite[e.g.][]{Karinen:2002}.   While it is agreed that anthropogenic power line harmonic radiation does exist in the magnetosphere, the magnitude of any effects of this radiation on the trapped electron populations remains uncertain.  

Another persistent anthropogenic radiation in the magnetosphere is that produced by the widespread distribution of VLF and RF transmitters around the world.   The radiation from these transmitters, used for navigation and communications, is known to disturb the trapped electron population of the magnetosphere.   In addition, over several decades, purposeful VLF transmissions in the form of controlled experiments have been conducted from spacecraft and from the ground (one of the more notable and long-lasting set of ground experiments was from Siple Station, beginning in the early 1970s and extending to the late 1980s).  These topics are addressed in Section~\ref{sec:VLF}.

In addition to purposeful short-term VLF radiation experiments to understand wave growth and trapped particle interactions, a number of other short-term experimental techniques have been developed and exploited over time.   The most dramatic of these were the explosion of nuclear bombs in the near-space environment, discussed in Section~\ref{sec:bombs}.   While these nuclear experiments were short term, their effects on the space environment in terms of enhanced radiation belt electrons persisted for weeks and months, forming artificial radiation belts (Section~\ref{subsec:belts}), with damaging radiation effects on flying spacecraft, as described in Section~\ref{sec:satellites}. The geophysical and geomagnetic effects of these explosions are reviewed in Section~\ref{sec:geomag}.

Other short-term experiments directed toward specific understanding goals and with limited lasting effects on the magnetosphere include the injection of clouds of barium into the ionosphere and lower altitude magnetosphere, and the heating of the ionosphere by high power RF transmitters.   Ionosphere RF heating experiments have had several objectives, including the generation of ULF waves with f $\lesssim$1 Hz.  Such waves, if propagated into the magnetosphere, could potentially interact with the gyro frequency of magnetosphere ions, as VLF waves do with electrons.    These topics are discussed inSection~\ref{sec:HF}.   

The exhaust from launches of large rockets and the controlled firings of the engines of the space shuttles were used specifically for modifications of the ionosphere -- the production of ionosphere ``holes'' \cite[e.g.][]{Mendillo:1981, Mendillo:1987, Meier:2011}.   Such holes can change the propagation conditions for ground-to-satellite (and the reverse) signals, and can in some instances enable the measurement of galactic cosmic radio noise at frequencies lower than normally possible under the conducting ionosphere.   

At times, other possibilities of exciting ULF waves from the ground into the magnetosphere have been discussed.    Generation possibilities have included large current loops on Earth's surface, vertical or horizontal dipoles, and driving a current in seawater around a peninsula \cite[]{Fraser-Smith:1981}.   \cite{Fraser-Smith:1978} reported ULF noise in ground-based measuring systems that is produced by the San Francisco BART (Bay Area Rapid Transit) system.   No evidence of such frequency noise propagating into the magnetosphere has been reported.    This BART-effect is similar to the long-recognized magnetic noise that is produced by trains passing in the vicinity of a magnetic field measuring site.  

The permanent existence, and growth, of power grids and of VLF transmitters around the globe means that it is unlikely that Earth's present-day space environment is entirely ``natural'' -- that is, that the environment today is the environment that existed at the onset of the 19th century.  This can be concluded even though there continue to exist major uncertainties as to the nature of the physical processes that operate under the influence of both the natural environment and the anthropogenically-produced waves.   As new techniques are considered for human modification of elements of Earth's space environment, it is important to carefully assess the short-term and long-term implications of anthropogenic modifications in order to arrive at final experiment design, and even decision to proceed.


\section{High-Altitude Nuclear Explosions}
\label{sec:bombs}

Missiles that were able to launch the Soviet Sputnik satellite could also deliver weapons across the world.    At the Livermore branch of the Lawrence Radiation Laboratory (now the Lawrence Livermore National Laboratory) where he was employed, Nicholas Christofilos proposed in October 1957 (just after the launch of Sputnik and months before the discovery of the Van Allen Radiation belts) a defensive way to intercept and destroy intercontinental missiles \cite[see pp 55-56 in][]{Finkbeiner:2006}.   He suggested that by exploding a nuclear bomb in the upper atmosphere, the electrons from the fission process would be trapped in Earth's magnetic field.  This huge cloud of trapped electrons would destroy an incoming missile.   

Christofilos's idea was followed up and supported by the Director of the Advanced Research project Agency (ARPA), Herbert York, as the Argus experiments in 1958 \cite[]{Christofilos:1959a, Christofilos:1959b}. The Argus shots demonstrated the feasibility of the concept and also showed that such injections of electrons could damage spacecraft flying through such a cloud: anthropogenic space weather. Thus, high-altitude explosions of nuclear devices could be offensive as well as Christofilos's original proposal of defense against missiles. On the other hand, the devastating electromagnetic pulse (EMP) that could be produced by the high altitude explosion of a nuclear device was only observed years later following the Starfish Prime event of July 9, 1962. A list of high-altitude nuclear explosions is given in \tablename~\ref{tab:tests}. 

\begin{table*}[h]
\begin{center}
\caption{List of high-altitude nuclear explosions \cite[]{Wikipedia:2016hemp}}
\label{tab:tests}       
\begin{tabular}{L{1.1in}L{0.45in}R{0.7in}S[table-format=6.1]S[table-format=5.2]}
\hline\noalign{\smallskip}
Designation & Country & Date\hspace{0.3in} & \text{\hspace{0.35in}Altitude} &  \text{\hspace{0.2in}Yield (kt)} \\
\noalign{\smallskip}\hline\noalign{\smallskip}
Yucca & USA & Apr 28, 1958 & 26 km & 1.7  \\
Teak & USA & Aug 1, 1958 &77 km & 3.8$\times10^3$ \\
Orange & USA & Aug 12, 1958 & 43 km & 3.8$\times10^3$ \\
Argus I & USA & Aug 27, 1958 & 200 km & 1.7 \\
Argus II & USA & Aug 30, 1958 & 240 km & 1.7  \\
Argus III & USA & Sep 6, 1958 & 540 km & 1.7 \\
Test\#88 & USSR & Sep 6, 1961 & 23 km & 10.5 \\
Test\#115 & USSR & Oct 6, 1961 & 41 km & 40  \\
Test\#127 & USSR & Oct 27, 1961 & 150 km & 1.2 \\
Test\#128 & USSR & Oct 27, 1961 & 300 km & 1.2 \\
Starfish Prime & USA & Jul 9, 1962 & 400 km & 1.4$\times10^3$ \\
Checkmate & USA & Oct 20, 1962 & 147 km & 7  \\
Test\#184 & USSR & Oct 22, 1962 & 290 km & 300 \\
Bluegill Triple Prime & USA & Oct 26, 1962 & 50 km & 410 \\
Test\#187 & USSR & Oct 28, 1962 & 150 km & 300 \\
Kingfish & USA & Nov 1, 1962 & 97 km & 410 \\
Test\#195 & USSR & Nov 1, 1962 & 59 km & 300 \\
\noalign{\smallskip}\hline
\end{tabular}
\end{center}
\end{table*}

The EMP generated by a high altitude nuclear explosion is one of a small number of threats that can hold our society at risk of catastrophic consequences. The increasingly pervasive use of electronics of all forms represents the greatest source of vulnerability to attack by EMP. When a nuclear explosion occurs at high altitude, the EMP signal it produces will cover the wide geographic region within the line of sight of the detonation. This broad band, high amplitude EMP, when coupled into sensitive electronics, has the capability to produce widespread and long lasting disruption and damage to the critical infrastructures that underpin the fabric of U.S. society.

\begin{figure*}[h]
\includegraphics[width=1\textwidth]{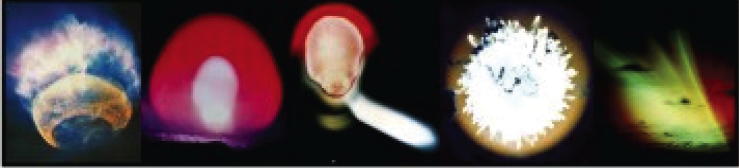}
\caption{From left to right, the Orange, Teak, Kingfish, Checkmate, and Starfish high-altitude
nuclear tests conducted in 1958 and 1962 by the United States near Johnston Island in the mid-Pacific 
\cite[from][]{Foster:2008}.}
\label{fig:bombs}
\end{figure*}

High-altitude nuclear explosions have vastly different EMP effects depending on the geomagnetic location and burst altitude. \figurename~\ref{fig:bombs} shows images of atmospheric nuclear bursts carried out by the U.S. over Johnston Island (geomagnetic latitude 10.5$^\circ$) during the 1958--1962 time period \cite[]{Foster:2008}.

From the very first nuclear test in July 1945 on, electromagnetic effects were always a threat for experimenters trying to record weapon performance data electronically. With bunkered and electrically shielded apparatus, data could be recorded successfully, but the power supplies used to convert AC current to direct current for the vacuum tubes would be burned out after the event.

Around 1960, questions were raised about possible damaging effects of EMP on the Minuteman missile system, which was in development at that time. This system was supposed to be able to launch retaliatory missiles after being subjected to nuclear attack. It was suggested that a large nuclear burst in the missile farms could expel the geomagnetic field from a large volume of air and ground, rapidly changing the magnetic flux. This would induce currents in cables that might be large enough to burn them out, preventing the retaliatory launches.

The Argus nuclear tests were the first experiment to explode nuclear weapons above the dense atmosphere. While these tests produced some interesting effects, the yield of the fission device (1.7 kt) was not large enough to produce electromagnetic effects over a wide area. 

\begin{figure}[h]
\floatbox[{\capbeside\thisfloatsetup{capbesideposition={left,top},
capbesidewidth=0.3\columnwidth}}]{figure}[\FBwidth]
{\caption{
The various phases of a generic HEMP signal \cite[from][]{Savage:2010}. For comparison, we also show a typical electromagnetic signal from lightning.
\label{fig:EMPphases}}}
{\includegraphics[width=0.65\textwidth]{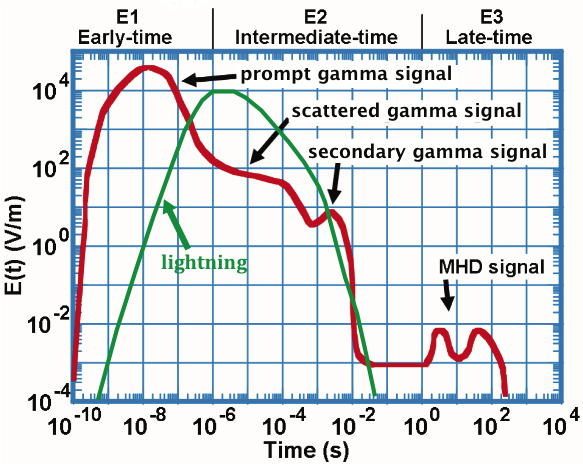}}
\end{figure}

High-altitude nuclear EMP (HEMP) is a complex multi-pulse phenomenon, usually described in terms of three components, ``E1'' (early phase), ``E2'' (intermediate phase) and ``E3'' (late phase, or MHD phase). A concise summary of these phases is shown in \figurename~\ref{fig:EMPphases}. Since the E2 phase is often compared to lightning the figure also shows a typical electromagnetic signal from lightning.

\subsection{E1 Phase}
\label{subsec:E1}

The true importance of the early phase of HEMP was revealed by the three high-altitude devices exploded at 400 km (Starfish), 97 km (Kingfish) and 50 km (Bluegill) altitudes that showed early electric field pulses of tens of kV/m peaking around 10 ns and lasting about a $\mu$s (see \figurename~\ref{fig:EMPphases}). The extremely short duration of the initial signal presented a puzzle that was solved by Conrad Longmire in 1964. Two decades later \cite{Longmire:1986} published an unclassified report that explains the fundamental physics of the early phase EMP (the E1 phase).

\begin{figure}[h]
\floatbox[{\capbeside\thisfloatsetup{capbesideposition={left,top},
capbesidewidth=0.4\columnwidth}}]{figure}[\FBwidth]
{\caption{
General basis of the E1 HEMP generation process. Gammas from the nuclear burst interact with the upper atmosphere generating Compton electrons, which are turned in the Earth's geomagnetic field, and produce a transverse current that radiates an EM pulse towards the Earth \cite[from][]{Savage:2010}.
\label{fig:E1generation} }}
{\includegraphics[width=0.55\textwidth]{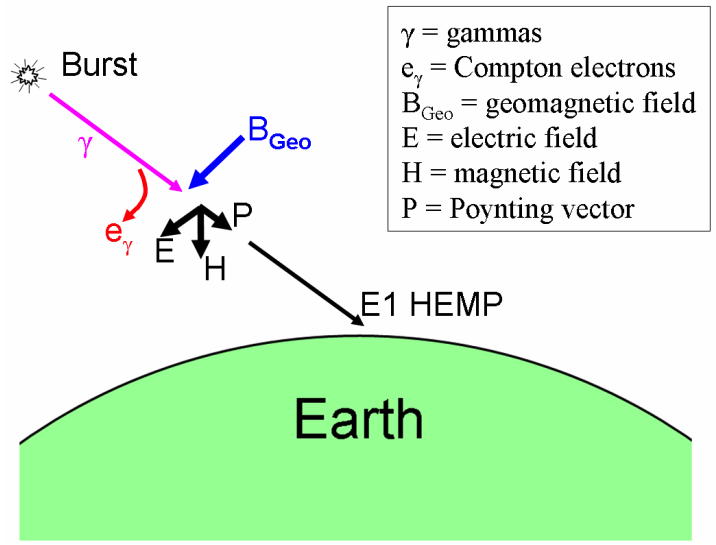}}
\end{figure}

\figurename~\ref{fig:E1generation} shows a general diagram of the E1 HEMP process. A nuclear burst puts out a fast pulse of gamma rays. This thin shell of photons streams outward, including downward toward the Earth and its exponentially increasing air density. Once low enough in altitude, the gammas start striking air molecules, knocking electrons off (which mostly move outward from the explosion). The Earth's magnetic field causes the electrons to turn coherently looping around the magnetic field, and this constitutes an electric current, which generates an EM signal, much like the currents on a transmitting loop antenna (magnetic dipole). This EM field propagates downward as an EM wave -- the E1 signal. This process was first explained by \cite{Longmire:1986}.

\cite{Longmire:1986} explains the basic physics of the E1 pulse in terms of an oversimplified ``flat Earth'' scenario. While this scenario is clearly unphysical, it clarifies the basic physics and explains the extremely short duration of the initial pulse. Here we briefly outline his derivation.

\subsubsection{Gammas and Compton electrons}

Nuclear bombs emit a small fraction, of the order of 0.3\%, of their energy in gamma rays. Thus a 1 Mt bomb (that has the explosive power of 1 Mt of TNT), produces a total energy of about $4.2\times10^{15}$ J and emits $\sim3$ kilotons or about $1.2\times10^{13}$ J in gamma rays with a typical average energy of 2 MeV ($3.2\times10^{-13}$ J). Thus the total number of gammas emitted by a 1 Mt bomb is of the order $N_\gamma \approx 4\times10^{25}$ gammas.

\begin{figure}[h]
\floatbox[{\capbeside\thisfloatsetup{capbesideposition={left,top},
capbesidewidth=0.4\columnwidth}}]{figure}[\FBwidth]
{\caption{
Schematic diagram of plane parallel E1 generation \cite[]{Longmire:1986}.
\label{fig:E1model} 
}}
{\includegraphics[width=0.4\textwidth]{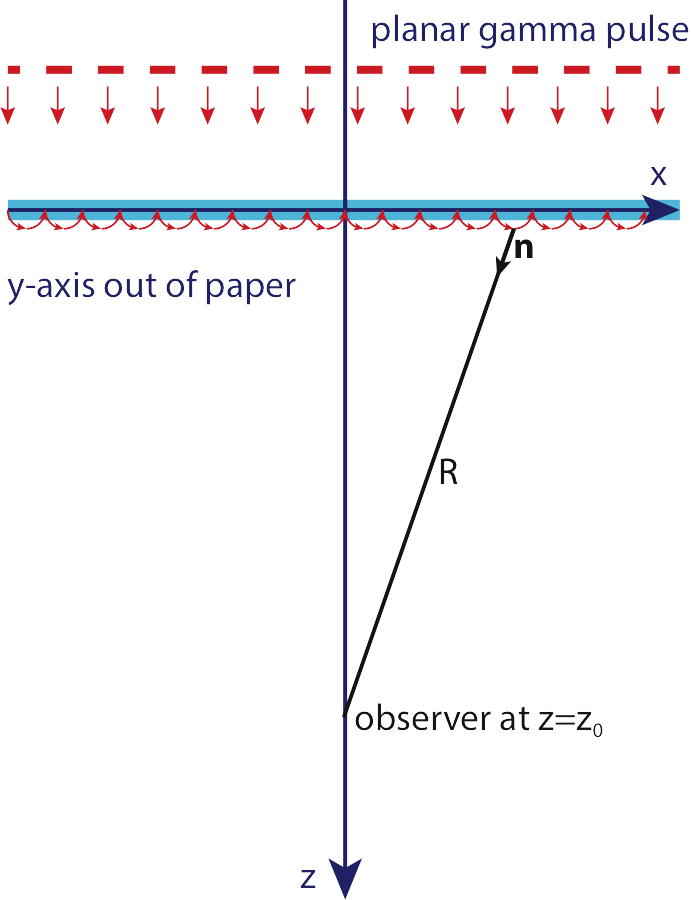}}
\end{figure}

The principal interaction of gamma rays with air molecules is Compton scattering. In this process, the gamma collides with an electron in the air molecule and knocks it out. In so doing, the gamma transfers part of its energy (on the average about half) to the electron, and is scattered into a new direction. The Compton recoil electron goes generally near the forward direction of the original gamma, thus a directed flux of gammas produces a directed electric current of Compton recoil electrons (see \figurename~\ref{fig:E1model}). 

A $\sim$2 MeV gamma penetrates an air column of $\sim$220 kg/m$^{2}$ before it is scattered. Assuming that the mass of the air column above the burst altitude is negligible (this was certainly true for the Starfish experiment) the gammas will have their first collision at an altitude above which the column density of air is $\sim$220 kg/m$^{2}$. This altitude is around $z_0=30$ km. Most gammas will suffer their first collision in a layer around $z_0$ with a thickness of about an atmospheric scale-height (at 30 km altitude the typical scale-height is $H_n\approx\!7$ km).

Assuming that gammas are uniformly emitted by the burst in every direction the total number of gammas crossing a unit cross sectional area at a distance $r$ from the burst is $N_{\scriptscriptstyle{\rm A}} = N_\gamma/4\pi r^2$. Assuming a 100 km burst altitude and 1 Mt bomb we get $N_{\scriptscriptstyle{\rm A}}\approx\!2.3\times10^{13}$ gammas/m$^2$. These gammas are scattered within an atmospheric scale-height around 30 km altitude,, so the number of Compton electrons produced in a unit volume is $n_\gamma\approx\!10^{11}$ electrons/m$^3$. The density of Compton electrons ($n_{{\scriptscriptstyle{\rm C}e}}$) would be the same as $n_\gamma$ if they did not move. Because they move in the same direction as the gammas with an average speed of about $0.94\,c$ (where $c$ is the speed of light), their actual density is about an order of magnitude larger than $n_\gamma$ (due to relativistic effects). In his estimates  \cite{Longmire:1986} used $n_{{\scriptscriptstyle{\rm C}e}}\approx\!10^{12}$ electrons/m$^3$. We note that if the Compton electrons all moved together at nearly the speed of light, they would make a current density of tens of A/m$^2$, that is a very substantial current density (about ten orders of magnitude larger than typical magnetospheric currents).

\subsubsection{Motion of Compton electrons}

For $\sim$2 MeV gammas the Compton recoil electrons have an average kinetic energy of about 1 MeV and their angular distribution strongly peaks in the forward direction. The velocity of a 1 MeV electron is about $0.94c$, while the relativistic mass is about three times larger than the rest mass. In his estimates \cite{Longmire:1986} used a magnetic field value of $B_0=5.6\times10^{-5}$ T that results in a gyroradius for 1 MeV electrons of $r_c=85$ m. The mean stopping range of the 1 MeV electrons due to collisions with air at 30 km altitude is $R_m\approx\!170$ m, taking into account the fact the the MeV electron gradually loses energy as it undergoes multiple collisions. In the process of stopping the MeV electron about $3\times10^4$ low energy electron-ion pairs (typical energies are $\sim\!10$ eV) are created. Because the velocities of secondary electrons are randomly distributed they generate no significant current. However, they create an electrically conducting layer that plays a role in the generation of the EMP pulse.

During its first gyration around the local magnetic field an average Compton electron would travel about $2\pi r_c =530$ m. However the stopping distance is only 170 m, so a Compton electron makes only a third of gyration before it is stopped. Substituting the numerical values used by \cite{Longmire:1986} we obtain a Compton current density of about $j_{\scriptscriptstyle{\rm C}}=-e c n_{{\scriptscriptstyle{\rm C}e}}\approx$ 50 A/m$^2$.

\subsubsection{Radiation}

Let us consider a planar, impulse function of gammas propagating through vacuum and arriving perpendicularly to a thin insulating sheet, as indicated in \figurename~\ref{fig:E1model}. We assume that electrons are knocked out of the sheet in the forward direction of the gammas, and that there is a magnetic field $B_0$ parallel to the sheet, in the y-direction. This field deflects the electrons in semicircles, until they return to the sheet and are stopped. An observer is located at a distance $z_0$ below the sheet, which is very large compared to the gyroradius of the electrons. These simplifications are introduced for ease of calculation and they still preserve the underlying physics.

At $t = 0$ the gamma pulse arrives at the thin sheet. The observer, however, will only get a start signal when the gammas arrive at his location, therefore the ``observing time'' (or delayed time) $T$ is related to the ``event time'' ($t$) at any value of $z$ as $T=t-z/c$. \cite{Longmire:1986} found the relation between event time and observing time for a radiating electron to be
\begin{equation}
   T = t - \frac{r_c}{c} \sin\omega_e t
\label{eq:time1}
\end{equation}
where $\omega_e$ is the electron gyrofrequency.
For short times ($\omega_e t \ll 1$) this expression becomes
\begin{equation}
   T = \left(1 - \frac{v_e}{c}\right) t
\label{eq:time2}
\end{equation}
This means that for MeV electrons the observing time initially advances nearly twenty times slower than the event time, resulting in an extremely short observed pulse.

\begin{figure}[h]
\floatbox[{\capbeside\thisfloatsetup{capbesideposition={left,top},
capbesidewidth=0.4\columnwidth}}]{figure}[\FBwidth]
{\caption{
he time-dependent factor in the field radiated by an electron with $v_e/c = 0.94$ that 
makes one-half turn in a magnetic field as observed in ``event time'' ($t$) and ``observer time'' ($T_0$). Due to relativistic effects the observer detects a $<$10 ns signal of positive $E_x$ \cite[from][]{Longmire:1986}.
\label{fig:Efield}
}}
{\includegraphics[width=0.55\textwidth]{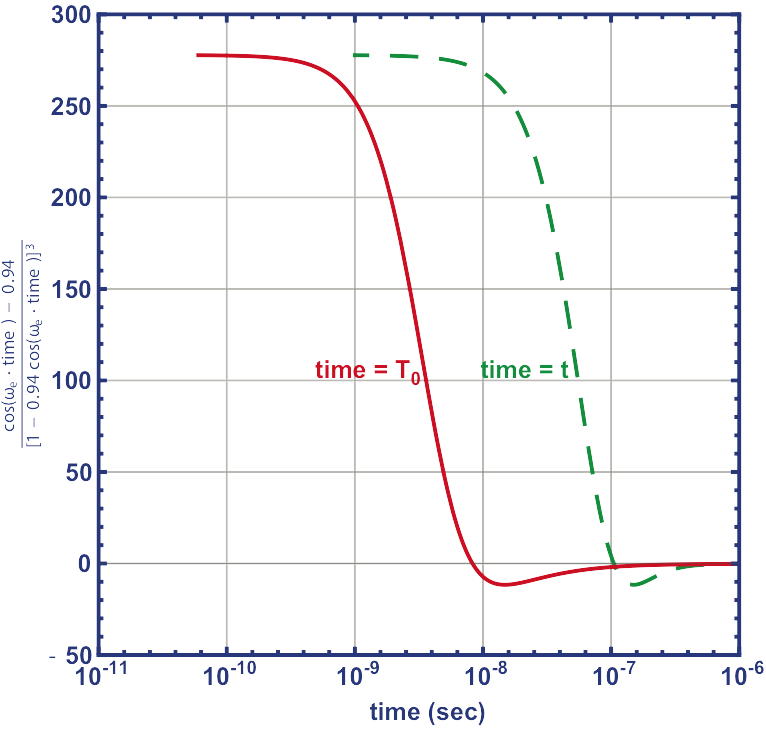}}
\end{figure}

\cite{Longmire:1986} also derived the radiated electric field associated with the motion of an electron that was born at $t=0$ at the origin of the coordinate system. The signal is detected by an observer located at a distance $z_0$ below the origin (see \figurename~\ref{fig:E1model}). In our simple model the horizontal electric field is
\begin{equation}
   E_x = E_0 \frac{\cos\omega_e t - \frac{v_e}{c}}{\left(1- \frac{v_e}{c}\cos\omega_e t\right)^3}
\label{eq:Efield}
\end{equation}
where $E_0$ is a time-independent constant. The time-dependent factor in $E_x$ is shown in \figurename~\ref{fig:Efield} as a function of both $T_0$ (i.e., as it would appear to the observer) and $t$ (as the electron experiences it). We note the very large amplitude at early times. The end of the electron trajectory after one-half turn in the magnetic field corresponds to $\omega_e T_0/\pi = 1$ or $T_0\approx1$ $\mu$s. No more field is radiated to our observer after this time (the radiation emitted in the starting and stopping of the electron, which may be called bremsstrahlung, vanishes in the z-direction) and the observed $E_x$ signal becomes small and negative after less than 10 ns. Integrating the electric field signal coming from electrons within a column of radius $\rho$ gives the total electric field seen by the observer \cite[]{Longmire:1986}:
\begin{equation}
   {\cal{E}}_x = \frac{1}{2}Z_0 e v_e N_{\scriptscriptstyle{\rm A}} \frac{\sin \omega_e t}{1 - \frac{v_e}{c}  \cos \omega_e t}
\label{eq:Etotal}
\end{equation}
where $N_{\scriptscriptstyle{\rm A}}$ is the column density of electrons in the thin layer where the gammas are absorbed. Since in this simple model every gamma creates an electron, $N_{\scriptscriptstyle{\rm A}}$ is the column density of $2$ MeV gammas hitting the atmosphere shortly after the burst. As we discussed earlier $N_{\scriptscriptstyle{\rm A}} \sim\!2.3 \times10^{13}$ m$^{-2}$ and thus the amplitude of the total electric field at the observer is 
\begin{equation}
   {\cal{E}} = \frac{1}{2}Z_0 e v_e N_{\scriptscriptstyle{\rm A}} \approx 2\times 10^5 \quad {\rm A/m}
\label{eq:Eamp}
\end{equation}

Below the source region, where the gammas have been mostly scattered, the HEMP propagates as a free wave without further buildup or attenuation. This is the signal that reaches the surface and can cause major disruptions in technological systems.

\subsubsection{Exposed Area}

\begin{figure}[b]
\begin{center}
\includegraphics[width=0.85\columnwidth]{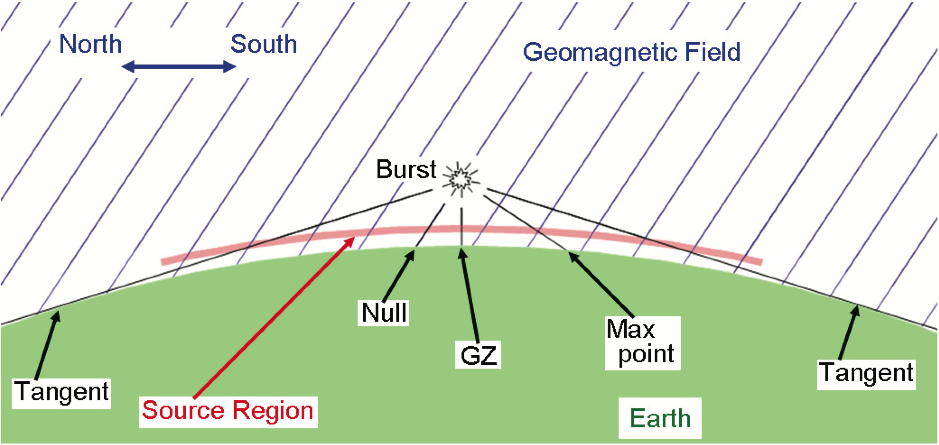}
\vspace{-0.25in}
\caption{
Typical geometry of E1 phase. There are three special points on the ground: Ground Zero (GZ) -- is the point directly below the burst, Null Point -- where the observer ray and geomagnetic field lines are parallel, and Max Point -- where the E1 HEMP has its maximum peak level \cite[from][]{Savage:2010}.
\label{fig:E1geometry} 
}
\end{center}
\end{figure}

The area impacted by HEMP E1 phase is controlled by the burst altitude and the the direction of the local geomagnetic field. \figurename~\ref{fig:E1geometry} shows three special points. The rays originating at the burst and that are tangent to the Earth's surface define the maximum extent of the HEMP E1 exposure region. Ground Zero is the point directly below the burst where the observer ray goes straight down. The red region in the atmosphere is the ``source region'' where most of the burst gammas interact with the atmosphere. The other two special positions in \figurename~\ref{fig:E1geometry} are related to the geomagnetic field. The ``null point'' is where the observer ray and geomagnetic field lines are parallel. For this point the E1 HEMP is very low (ideally it would be zero). On the opposite side of Ground Zero is the ``max field point''. This is the location where the HEMP E1 has its maximum peak level (where the observer ray is perpendicular to the geomagnetic field lines). This point could be called the ``geomagnetic max point''. 

\begin{figure}
\begin{center}
\includegraphics[width=0.85\columnwidth]{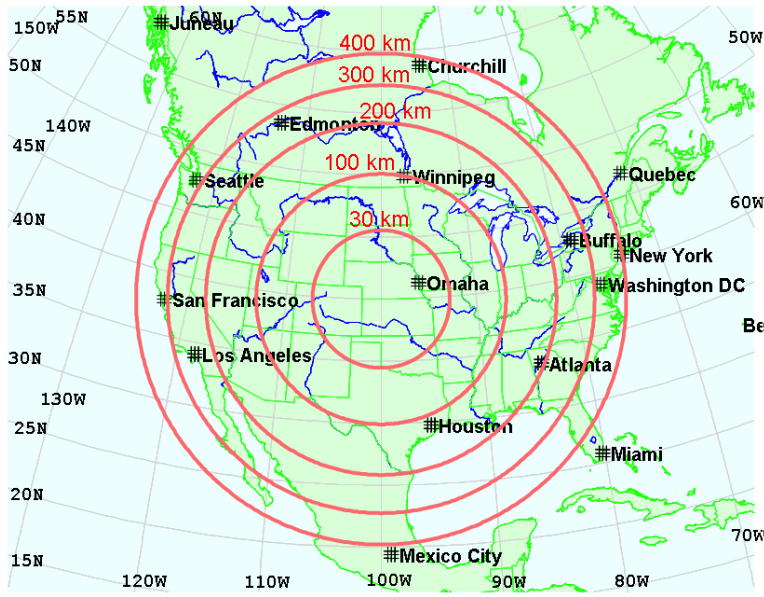}
\vspace{-0.25in}
\caption{
Samples of E1 HEMP exposed regions for several burst heights. The red circles show the exposed
regions for the given burst heights, for a nuclear burst over the central U.S. \cite[from][]{Savage:2010}.
\label{fig:E1impact} 
}
\end{center}
\end{figure}

\figurename~\ref{fig:E1impact} shows the area coverage for a HEMP over the U.S., for several burst altitudes. One can see that a nuclear explosion at an altitude of $\sim$400 km above Oklahoma would impact the entire continental United States.

\subsection{E2 Phase}
\label{subsec:E2}

As we discussed above the E1 phase of the HEMP is associated with unscattered gammas directly coming from the nuclear explosion. However, as we discussed in Section~\ref{subsec:E1} the direct gammas travel through $\sim$220 kg/m$^{2}$ of air before they hit the E1 source layer. About a fraction of $1/e\approx37\%$ of the $2$ MeV gammas interact with air molecules before reaching the E1 source layer resulting in a short time delay in the arrival time of these scattered gammas to the source layer. These variously-produced gammas lead to an impulsive Compton electron current (due to the separation of the electrons from their parent molecules) that depends on the polar angle because of the atmospheric density gradient. There is also a non-compensated vertical current directly below the explosion. 

The end result of these secondary gammas is that at the end they also produce an electromagnetic pulse similar to the E1 phase. While the basic physics of the generation of this secondary signal is basically the same as we discussed in Section~\ref{subsec:E1}, there is one big difference: the secondary E2 signal lasts much longer (actually several thousand time longer) than the E1 signal. Consequently the peak of the E2 signal is several hundred times weaker than the peak of the E1 signal (see \figurename~\ref{fig:EMPphases}). Even though most of the E2 signal takes place within about 10 ms after the burst, the E2 phase only ends at about 1 s after the burst.

The E2 phase covers roughly the same geographic area as the E1 component and is similar to lightning in its time-dependence, but is far more geographically widespread in its character and somewhat lower in amplitude. In general, it would not be an issue for critical infrastructure systems since they have existing protective measures for defense against occasional lightning strikes. The most significant risk is synergistic, because the E2 component follows a small fraction of a second after the devastating E1 impact, which has the ability to impair or destroy many protective and control features. Thus the energy associated with the second component thus may be allowed to pass into and damage systems.

\subsection{E3 Phase}
\label{subsec:E3}

The E3 component is very different from E1 and E2. E3 is a very slow pulse, lasting tens to hundreds of seconds  (see \figurename~\ref{fig:EMPphases}). It is produced by two different physical mechanisms, both of which are associated with the continuum behavior of the medium. This is the reason why the E3 phase is also called the MHD (magnetohydrodynamic) phase of the HEMP.

The first phase that typically lasts about 1 to 10 seconds is called the E3A or ``Blast Wave'' phase. This is characterized by the explosive expansion of the fireball containing a large mass of ionized material. The expanding plasma cloud expels the geomagnetic field and creates a diamagnetic bubble. The distortion of the geomagnetic field generates a transient current system that creates the first MHD peak shown in \figurename~\ref{fig:EMPphases} \cite[cf.][]{Gilbert:2010}.

The second phase of E3, called E3B, takes place between about 10 and 300 seconds. It is created when the hot, ionized, diamagnetic debris bubble buoyantly rises in the upper atmosphere. As this conducting patch crosses geomagnetic field lines it generates currents that flow in the patch and distorts the ground magnetic fields on the surface. This late phase is also called ``Heave'' and it creates the second MHD peak shown in \figurename~\ref{fig:EMPphases} \cite[cf.][]{Gilbert:2010}. 

\subsubsection{Blast Wave}

About 75\% of the energy of a high-altitude nuclear explosion is emitted in the form of x-rays with an average energy of a few keV. About half of these x-rays are emitted downward and are absorbed in the atmosphere in the altitude range of 80 and 110 km. The primary x-rays interact with K-shell electrons in neutral molecules and knock them out producing primary photoelectrons. The missing K-shell electron causes secondary x-ray emission when an electron jumps from a high-shell orbit to fill the gap on the K-shell. The primary photoelectrons create an additional ionization cascade resulting in a high level of ionization and heating in the 80--110 km layer. The horizontal extent of this layer is similar to the E1 phase, since both the gamma rays and x-rays expand radially from the blast location. This hot, ionized layer acts to shield the ground from direct electromagnetic signals generated in the burst region and ``anchors'' the geomagnetic field lines.

\begin{figure}[h]
\floatbox[{\capbeside\thisfloatsetup{capbesideposition={left,top},
capbesidewidth=0.3\columnwidth}}]{figure}[\FBwidth]
{\caption{
Schematic of E3A Blast Wave phenomenology. The x-ray patch is shown at less that its actual
radius for illustrative purposes. \cite[from][]{Gilbert:2010}.
\label{fig:fireball} 
}}
{\includegraphics[width=0.65\textwidth]{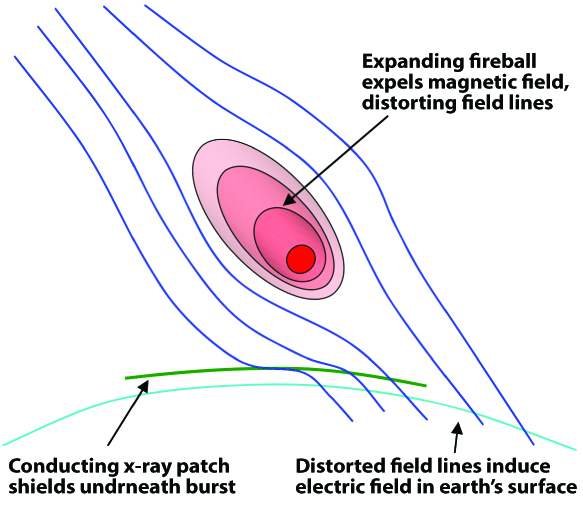}}
\end{figure}

About a quarter of the blast energy is kinetic energy of the weapon debris, which has become highly ionized and therefore has a high electrical conductivity. The debris expands outward, pushing the geomagnetic field out of the conducting region, and this forms a diamagnetic cavity or ``magnetic bubble.'' The initial expansion of the bubble is determined by the velocity of the debris, which is greater than 1000 km/s. Later expansion of the bubble depends on the blast altitude. For altitudes below about 300 km, the dominant effect slowing the expansion is the outside atmospheric pressure, and the bubble becomes asymmetric as it expands more easily upward into more rarefied air than downward into denser air. For higher blast altitudes, the atmospheric pressure is negligible and the expansion is slowed by the anisotropic magnetic pressure gradient force. At the end the kinetic energy of the debris is converted into magnetic energy of distorted field lines. Due to the anisotropy of the magnetic pressure the bubble expands more rapidly along the geomagnetic field lines and less rapidly in the perpendicular direction \cite[]{Karzas:1962}. This situation is depicted in \figurename~\ref{fig:fireball}. For observers at large distances from the burst, the perturbation of the magnetic field looks like a field-aligned dipole at the burst point. The resulting electric and magnetic fields are low, but they can exist over long time periods and large areas.

\subsubsection{Heave Phase}

\begin{figure}[h]
\begin{center}
\includegraphics[width=0.9\columnwidth]{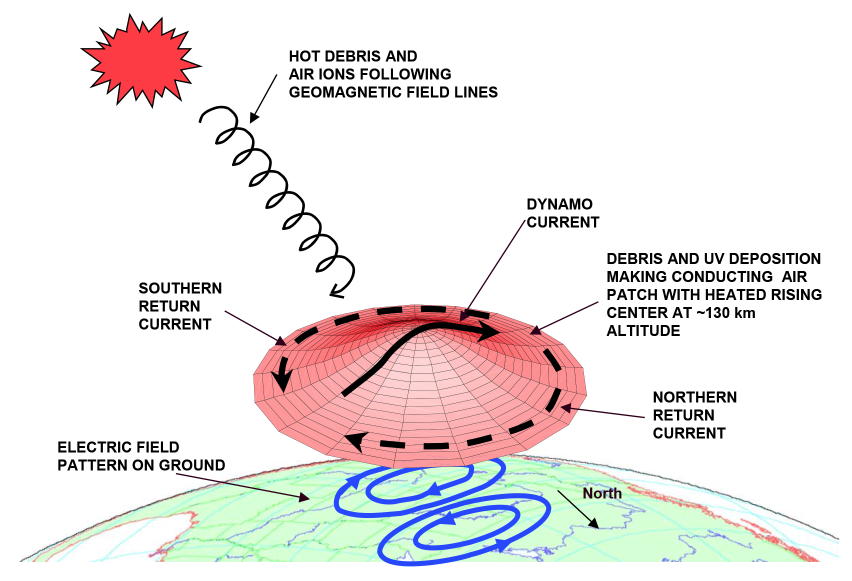}
\caption{Schematic of E3B Heave phenomenology \cite[from][]{Gilbert:2010}.
\label{fig:E3B}
}
\end{center}
\end{figure}

For bursts above $\sim$150 km altitude the second part of the EMP E3 phase is the  Heave effect (also called the E3B phase). The bomb debris and shock heated air ions stream downward along geomagnetic field lines until they deposit their energy at altitudes near 130 km. These two processes contribute to both the heating and additional ionization of the ionospheric E-layer. There is also heating of this region by UV radiation from the burst, and this heating is centered beneath the burst. After an initial brief period of expansion of the heated air, it buoyantly rises. The basic phenomenology of the heave generation is shown in \figurename~\ref{fig:E3B}.  The region with enhanced conductivity is depicted as a reddish ``hat'' with the heated center rising more rapidly than the brim. As this conducting layer rises across the geomagnetic field, a current is induced by the dynamo effect, as indicated by the solid arrow on the hat in \figurename~\ref{fig:E3B}. The dynamo current flowing to the west is accompanied by northern and southern return currents in regions where the heating and buoyant rise is smaller. This current system induces an oppositely oriented  ``two-cell'' current system in the ground, and the finite conductivity of the ground means that the current in the ground is accompanied by an electric field in the same direction, as shown by the blue pattern \cite[]{Gilbert:2010}.

The maximum electric field strength can be found in the region beneath the most highly heated portion of the atmosphere. Because the heating of the atmosphere is relatively localized (several hundred kilometers), the heave phase impacts a much smaller area than the blast wave.The electric field pattern mirrors the current
flow in the conducting heave region. The centers of the ``two-cell' patterns themselves are null field regions, where there is large cancellation between dynamo current flow and the return current flow.


\section{Artificial radiation belts}
\label{subsec:belts}

The first major scientific discovery of the Space Age was that the Earth is enshrouded in toroids, or ``belts,'' of very high-energy magnetically trapped charged particles \cite[]{VanAllen:1958, VanAllen:1959a}. Early observations of the radiation environment suggested that the Van Allen belts could be delineated into an inner zone dominated by high-energy protons and an outer zone dominated by high-energy electrons \cite[]{VanAllen:1959b}. Subsequent studies showed that electrons in the energy range 100 keV$<$E$<$1 MeV often populated both the inner and outer zones with a pronounced ``slot'' region relatively devoid of energetic electrons existing between them. The energy distribution, spatial extent and particle species makeup of the Van Allen belts has been subsequently explored by several space missions.

Experience has shown that the near-Earth space environment can cause significant operational anomalies and even spacecraft failures under certain circumstances \cite[e.g.][]{Fennell:2001}. The primary sources of spacecraft operational problems include energetic ions causing so-called single-event upsets (SEUs), moderate-energy electrons producing surface differential charging effects, and high-energy radiation belt electrons inducing deep-di\-e\-lec\-tric charging conditions \cite[see][]{Baker:2002}. SEU effects can be due to galactic cosmic rays, solar energetic particles, or trapped ions in Earth's inner Van Allen radiation belt. Surface charging is associated most closely with $\sim$$10$ to $\sim$$100$ keV energetic electrons during geomagnetically active times when spacecraft surfaces are in shadowed regions (or the entire satellite is in solar eclipse). Deep dielectric charging occurs most prominently when electrons of hundreds of keV to multiple-MeV energy are enhanced within the outer Van Allen radiation belt.

The natural space environment of the Earth has been extensively explored through the past six decades.  The Radiation Belt Storm Probes (Van Allen Probes) mission of NASA \cite[]{Mauk:2013}, launched on August 30, 2012, are currently bringing exceptionally detailed observations of the near-Earth space environment and its dynamic properties.  The salient properties of the multi-MeV electron environment of the Earth, as reported in a recent study \cite[]{Baker:2016}, include: (1) The outer Van Allen zone during strong geomagnetic storms can vary in absolute intensity by some six orders of magnitude on time scales of less than one day; (2) The magnetosphere is fully capable of accelerating relativistic electrons up to energies approaching 10 MeV on time scales of a few hours; (3) The radiation belt boundaries are determined both by natural and human-induced wave effects on the electron spatial and spectral properties (see below); and (4) The inner Van Allen radiation zone (1.1$\lesssim$L$\lesssim$2.5) has been essentially devoid of multi-MeV electrons for the entire Van Allen Probes era (2012-present).

\begin{figure}[h]
\floatbox[{\capbeside\thisfloatsetup{capbesideposition={left,top},
capbesidewidth=0.3\columnwidth}}]{figure}[\FBwidth]
{\caption{
Plots of passes by the Explorer IV satellite through the enhanced energetic electron fluxes in the artificial radiation belts created by the Argus nuclear explosions. (a) A pass through the Argus II belt, with the decaying Argus I belt still visible on 31 August 1958.   (b) A pass through the Argus III artificial belt on September 6, 1958. \cite[From][]{VanAllen:1959c}
\label{fig:andre1}
}}
{\includegraphics[width=0.65\textwidth]{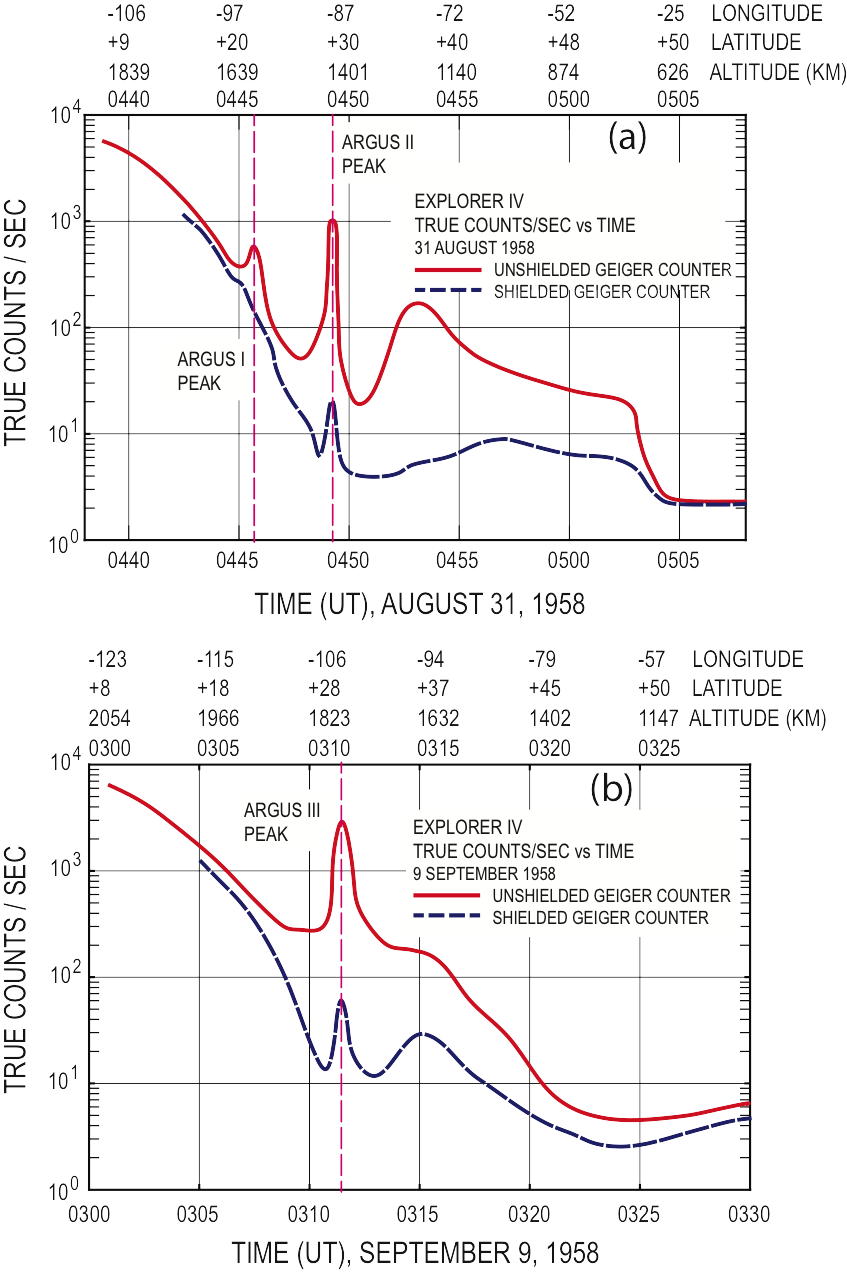}}
\end{figure}

As introduced in Section~\ref{sec:bombs}, secret activities were underway in 1957--58 to explode nuclear weapons in space to create artificial radiation regions in Earth?s magnetic field  \cite[]{Christofilos:1959a, Christofilos:1959b}.  Christofilos envisioned two physical effects: (1) A strong, prompt enhancement of ionospheric ionization in the vicinity of the nuclear explosion, thereby producing strong disruption of radio communication at VHF frequencies; and (2) Strong injection and trapping of high-energy (E $>$1 MeV) electrons in Earth's dipole-like magnetic field.  A key question in Christofilos' work was what fraction of emitted fission electrons and positrons from the nuclear explosion would become trapped in in the Earth's field  \cite[see][]{VanAllen:1959c, Walt:1997}.

\begin{figure}[h]
\floatbox[{\capbeside\thisfloatsetup{capbesideposition={left,top},
capbesidewidth=0.3\columnwidth}}]{figure}[\FBwidth]
{\caption{
The decay of energetic electron fluxes in the artificial radiation belt created by the Argus III nuclear explosion, observed by the Explorer IV satellite over several weeks.  The plot shows the product of the maximum true counting rate at the center of the Argus III shell with the geometric width of the shell at half maximum, to provide a measure of the electron content of the belt and its decrease with time.  \cite[From][]{VanAllen:1959c}
\label{fig:andre2}
}}
{\includegraphics[width=0.65\textwidth]{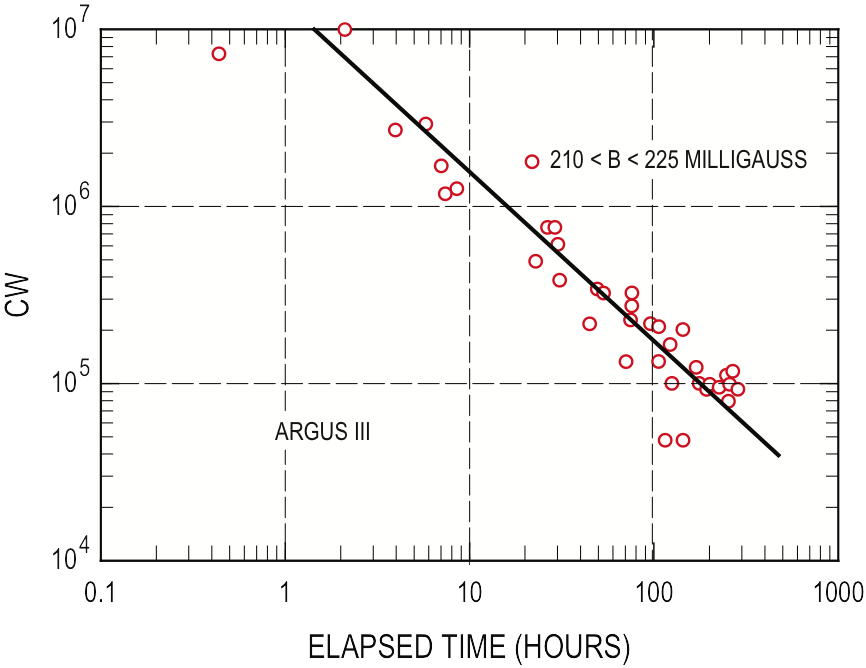}}
\end{figure}

After the successful deployment and initial operation of the Explorer I and III missions, Van Allen and coworkers were brought into the secret program to test the Christofilos effect \cite[]{VanAllen:1997}. In anticipation of high-altitude nuclear tests, starting in May 1958, the Explorer IV spacecraft was instrumented by Van Allen and his team in record time and was launched on July 26, 1958 \cite[]{VanAllen:1959e}. As seen in Table~\ref{tab:tests} above, there had already been by that time, one relatively small (1.7 kiloton) atmospheric test (codenamed Yucca) on April 28, 1958. In August 1958 there were large weapon tests of about 4 Mt yields (Teak and Orange), but these were at relatively low heights (some tens of km altitude) and there were only small effects seen in the radiation belts \cite[]{VanAllen:1997}. The truly high-altitude weapons tests were in Operation Argus \cite[]{Jones:1982} and, as shown in Table~\ref{tab:tests}, these were small weapons at several hundred kilometers altitude. As reported \cite[]{VanAllen:1959c, Brown:1966, VanAllen:1997, Walt:1997} the Argus tests produced discernible, but not major, perturbations of the radiation environment. All three artificial radiation shells were clearly observable by the Explorer IV instruments designed for the purpose (see \figurename~\ref{fig:andre1}). \cite{VanAllen:1997} summarized that Argus I and II each produced shells of artificially injected electrons that had a lifetime of about 3 weeks. The explosions, on August 27 and 30, 1962 were followed by major -- natural -- geomagnetic storms.  It is likely that these shortened the lifetime of the artificial radiation belts, as the loss of energetic trapped electron due to geomagnetic activity is well documented \cite[e.g.][]{Horne:2009a}.  The Argus III shot also produced a shell of electrons.  Argus III had a lifetime of about a month, see the decay of the observed peak intensities in \figurename~\ref{fig:andre2}.   All of these nuclear weapon-injected electrons had -- as measured by Explorer IV sensors -- a very different energy spectrum than the naturally occurring radiation belt population. Thus, observations could readily distinguish fission electrons from ``normal'' Van Allen belt particles \cite[]{VanAllen:1959c}.

Neither the Explorer IV satellite, nor the Argus experiments were a part of the contribution of the United States to the International Geophysical Year (IGY) that ran from July 1957 to December 1958 during which over 60 countries collaborated on investigations of a range of geophysical phenomena.  But Van Allen, in a letter dated February 21, 1959, requested the declassification of the geophysically relevant results of both the Argus experiments and the observations made by Explorer IV.  In the letter, Van Allen wrote that the Argus experiments ``undoubtedly constitute the greatest geophysical experiment ever conducted by man.'' The earliest published record, describing both geomagnetic effects and an artificial aurora, identified the nuclear tests as the cause of these geophysical phenomena \cite[]{Cullington:1958}. Following Van Allen's letter and the declassification of the Argus tests, the Explorer IV observations were presented at a Symposium of the US National Academy of Sciences by \cite{VanAllen:1959d}. Observations of the trapped electrons from the Argus tests were also made by a series of 19 dedicated sounding rockets \cite[]{Allen:1959} who noted that the energy spectrum of the electrons was a modified form of the fission spectrum, with a lower than expected high energy contribution, confirming the observations mentioned above by Explorer IV.

\begin{figure}[h]
\floatbox[{\capbeside\thisfloatsetup{capbesideposition={left,top},
capbesidewidth=0.3\columnwidth}}]{figure}[\FBwidth]
{\caption{
The spatial distribution of high energy electron fluxes measured by the Injun I satellite in the newly formed artificial radiation belt in the hours following the explosion on July 9, 1962. \cite[From][]{VanAllen:1963, VanAllen:1966}.
\label{fig:andre3}
}}
{\includegraphics[width=0.65\textwidth]{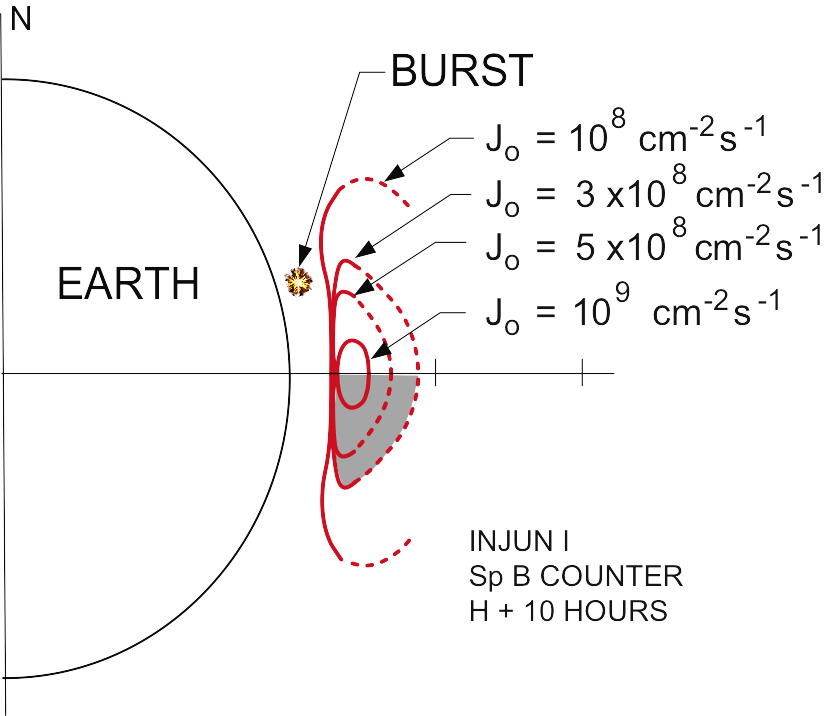}}
\end{figure}

The series of high-altitude nuclear detonations in 1962 by both the United States and the Soviet Union (see Table~\ref{tab:tests}) provided a greater opportunity to observe and analyze the formation of artificial radiation belts.  Their effects on satellites and on geophysical phenomena are discussed below in Sections \ref{sec:satellites} and \ref{sec:geomag}.  Here we describe the formation and evolution of the radiation belts formed from energetic electrons originating from the decay of nuclear fission products.  

The best documented explosion, with the longest lasting effects, was the Starfish burst on July 9, 1962 at 400 km altitude over Johnston Island in the Pacific, with a yield of about 1.4 Mt.  From classified observations at the time that became public only much later \cite[]{Dyal:2006}, it is now known that in the diamagnetic cavity that was formed at the time of the explosion the flux of energetic beta particles (electrons and positrons) was approximately uniform through the volume, with intensities of about $3 \times 10^{11}$ particles cm$^{-2}$ s$^{-1}$, for a duration of at least 7 seconds.  At 34 s after the explosion, the flux injected into the geomagnetic field from the diamagnetic cavity was $2.5 \times 10^{10}$ beta cm$^{-2}$ s$^{-1}$.  Observations by two of the rocket flights above the detonation site that were reported by \cite{Dyal:2006} were made at a location (in McIlwain's B-L coordinate system) at B = 0.153 gauss and L = 1.269.  The fluxes measured were $4 \times 10^{10}$ beta cm$^{-2}$ s$^{-1}$ at 25 seconds after the explosion and $6 \times 10^9$ beta cm$^{-2}$ s$^{-1}$ at 10 minutes after the explosion. 

It is this population of energetic electrons that were trapped and entrained by the Earth's magnetic field and formed the artificial radiation belt.  The first observations of the Starfish explosion were made by the UK-US Ariel-1 satellite \cite[]{Durney:1962}), the Soviet Kosmos-5 satellite \cite[]{Galperin:1964} and by the University of Iowa's Injun I satellite \cite[]{OBrien:1962a, VanAllen:1963}, as described in more detail in Section~\ref{sec:satellites}.  In the hours that followed the detonation, Injun I successfully mapped the core of the newly created artificial belt of high energy electrons, as shown in \figurename~\ref{fig:andre3} [taken from \cite{VanAllen:1966}]. The peak flux in the central volume of the belt, shortly after the explosion, was $>$$10^9$ electrons cm$^{-2}$ s$^{-1}$, well matched to the estimated injected population calculated by \cite{Dyal:2006}.

\begin{figure}[h]
\floatbox[{\capbeside\thisfloatsetup{capbesideposition={left,top},
capbesidewidth=0.3\columnwidth}}]{figure}[\FBwidth]
{\caption{
Contours of omnidirectional counting rates (OCRs), electron channel 3 by Telstar-1, on Days 193-197, 1962, starting three days following the Starfish explosion.  OCR is measured in counts/sec; the  electron channel 3 responds to electrons of energy greater than about 400 keV, with maximum efficiency at about 500 keV.  The plot is in the R-$\lambda$ coordinate system, related to the McIlwain's B-L coordinate system \cite[see][]{Roberts:1964}.  \cite[From][]{Brown:1963a}
\label{fig:andre4}
}}
{\includegraphics[width=0.65\textwidth]{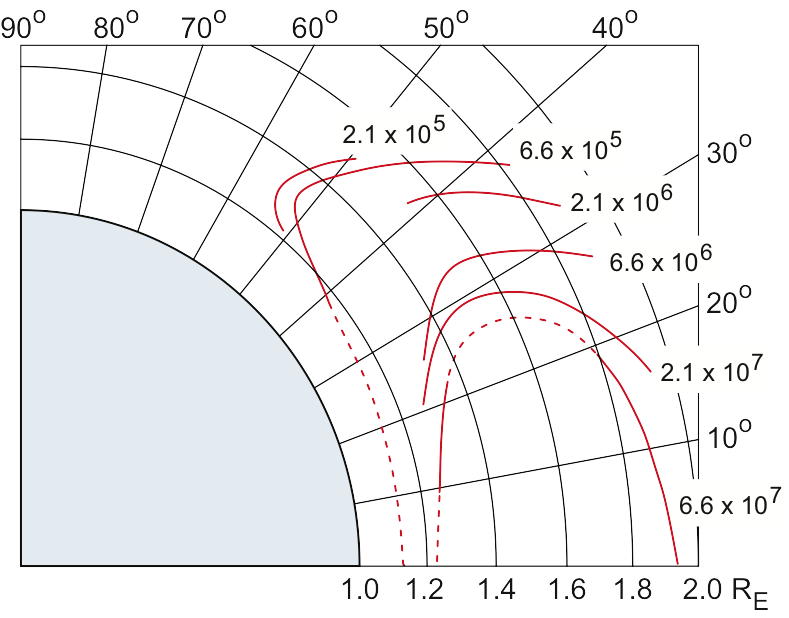}}
\end{figure}

The launch of Telstar-1, on the day following the Starfish explosion, brought a very powerful observation platform into near-earth space, extremely well instrumented (for the time) to monitor the energetic particle population of the natural as well as of the artificial radiation belts \cite[]{Brown:1963a}.  Helped by both its orbit (perigee = 952 km, apogee = 5632 km, inclination = 45$^\circ$) and its instrumentation, it was able to map the Starfish radiation belt and follow its evolution.  \figurename~\ref{fig:andre4} shows the distribution of energetic electron fluxes within the first week following the explosion.  Two aspects need to be noted.  First, the contours do not show the apparently much higher intensity core of the artificial belt that the Injun I observations implied.  This led at the time to a possible ambiguity between the two sets of measurements.  However, the Telstar-1 data do not contradict the existence of a higher intensity core, but represent simply different results due to the different energy response of the instruments and the different spatial sampling.  The second point to note is that the artificial radiation belt, at least initially, had a much greater extent in volume than implied by the explosion site.  The extent of the diamagnetic cavity, already described and the implication of the vertical expansion, driven by jets and driving a shock wave to a height of 1000 km or higher \cite[]{Colgate:1965}, can explain the population of L shells up to and beyond L = 2.

\begin{figure}[h]
\floatbox[{\capbeside\thisfloatsetup{capbesideposition={left,top},
capbesidewidth=0.3\columnwidth}}]{figure}[\FBwidth]
{\caption{
The omnidirectional counting rate between the launch of Telstar-1 and the first of the Soviet high altitude blasts on day 295.  The curves show the counting rates in a narrow ? ranges on individual L shells, close to the equator for L = 1.7 and increasingly larger l values for the higher L shells.  The clustering of observational points is due to the precession of the Telstar-1 orbit.  \cite[From][]{Brown:1966}
\label{fig:andre5}
}}
{\includegraphics[width=0.65\textwidth]{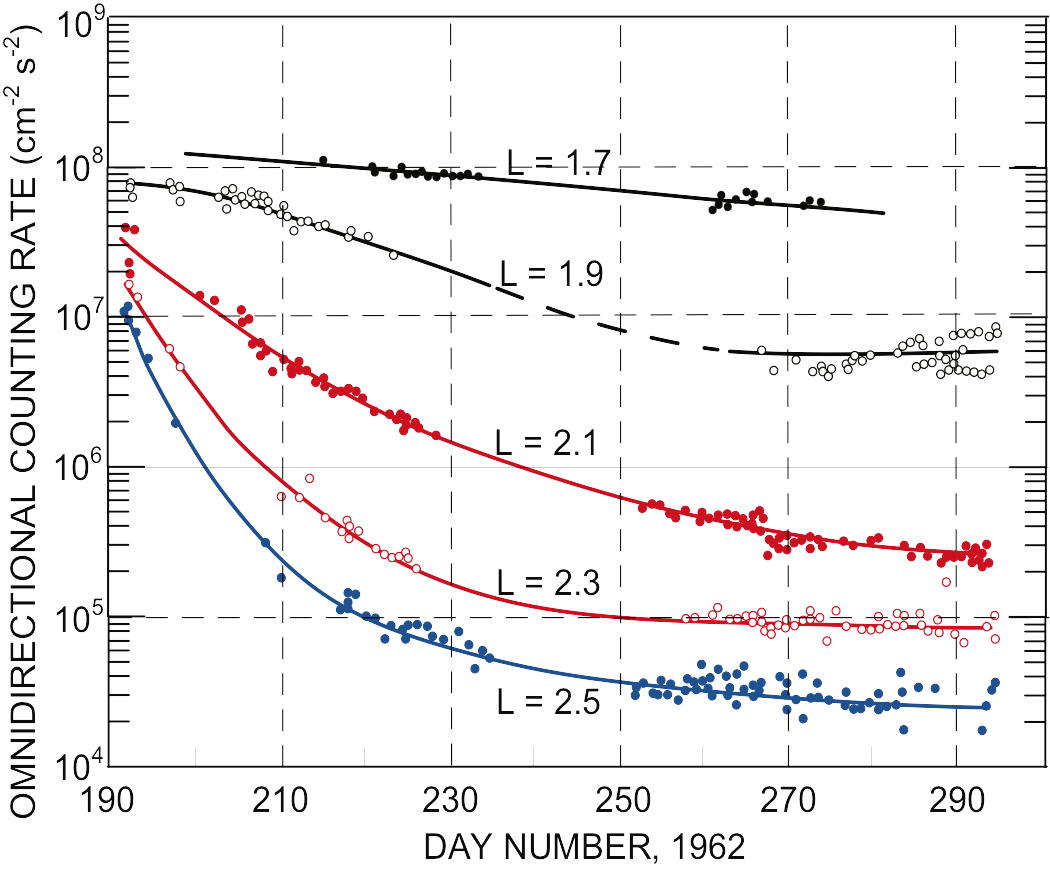}}
\end{figure}

Given the quasi-instantaneous injection of the population at the time of the explosion, the decay of electron fluxes in the artificial belts provided an opportunity to study the possible loss mechanism of these particles. \figurename~\ref{fig:andre5} follows the decay, as a function of L, of energetic electrons observed by Telstar-1 \cite[from][]{Brown:1966}.  The decay is faster as L increases towards the radiation belt slot region where electrons are lost mainly to pitch-angle scattering due to wave-particle interactions.  However, the artificial belt is quite stable at L = 1.7.  At lower L-values (L $<$1.2), atmospheric interactions  caused the decay of the fluxes \cite[]{Walt:1964}.  But at intermediate L-values, 1.3 $<$ L $<$ 2, the loss mechanism is not fully understood, and the lifetimes in the artificial belt were found to be several years.

\begin{figure}[h]
\floatbox[{\capbeside\thisfloatsetup{capbesideposition={left,top},
capbesidewidth=0.3\columnwidth}}]{figure}[\FBwidth]
{\caption{
Observations of the time evolution of energetic electron fluxes by Telstar-1, showing the decay of the fluxes following the Starfish nuclear test on 9 July 1962 and showing the temporary replenishment by the Soviet nuclear tests on 22 and 28 October and 1 November 1962. In the figure $x=\sqrt{1-B_0/B}$. \cite[From][]{Brown:1966}
\label{fig:andre6}
}}
{\includegraphics[width=0.65\textwidth]{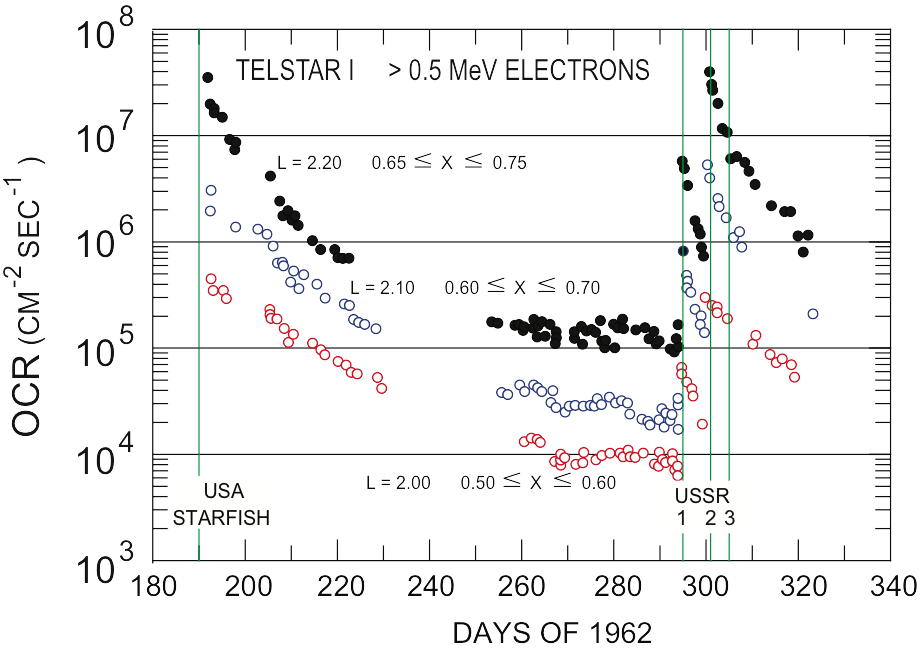}}
\end{figure}

The three Soviet nuclear explosions in late October and on November 1, 1962 replenished the artificial radiation belt, although apparently at somewhat higher L-values (likely to be due to the more northerly location of the explosions).  The new population, superimposed on the decaying Starfish belt, is well illustrated in \figurename~\ref{fig:andre6} \cite[from][]{Brown:1966}.  The peak intensities appeared close to that from Starfish; the decay constants appeared to be quite comparable to those experienced by the Starfish fluxes. 

\begin{figure}[h]
\floatbox[{\capbeside\thisfloatsetup{capbesideposition={left,top},
capbesidewidth=0.3\columnwidth}}]{figure}[\FBwidth]
{\caption{
The long-lasting energetic electron fluxes at L = 1.25, measured first by the Injun-I satellite in the aftermath of the Starfish detonation and, from its launch on December 13, 1962, by the Injun-III satellite.  After the initial fast decay, fluxes in the artificial radiation belt stabilized and decayed over many months.  Similar instruments on the two satellites responded to electrons between 1.5 and 5 MeV. \cite[From][]{VanAllen:1964}
\label{fig:andre7}
}}
{\includegraphics[width=0.65\textwidth]{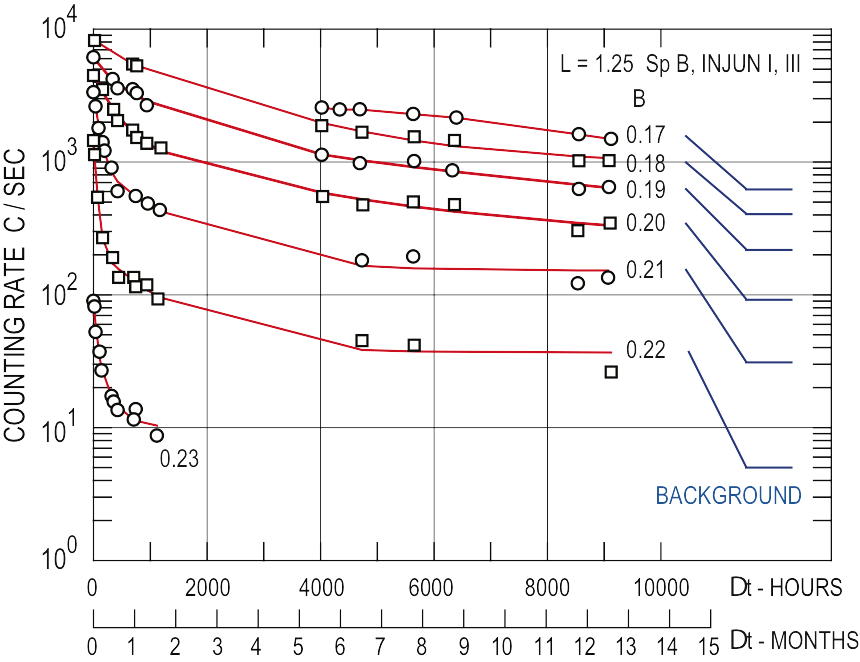}}
\end{figure}

The longer term stability at L = 1.25 was measured by comparing the fluxes observed by Injun I in the weeks after Starfish and those measured by the Injun III satellite (launched on December 13, 1962) by similar instruments.  The results are shown in \figurename~\ref{fig:andre7}.  It can be seen that following a fast initial decay, the near-equatorial Starfish artificial belt stabilized and the fluxes remained up to an order of magnitude higher for well over a year than the pre-Starfish natural fluxes. It is unclear, but likely, that the fluxes observed by Injun III were exclusively due to Starfish and were not influenced by any replenishment from the Soviet explosions that occurred about six weeks before The Injun III launch, because the core high intensities of the Soviet artificial belts were at higher L-values than the Starfish belt.

\begin{figure}[b]
\begin{center}
{\caption{
The peak of energetic electron fluxes (E $>$ 1.9 MeV) due to the third U.S.S.R. test on November 1, 1962 were distributed in a narrow range of L-shells around L = 1.765 at four epochs following the nuclear blast.  The height of the peak is decreasing with a time constant of days.  It is also broadening, due to cross-L diffusion.  The fluxes at lower L-values are due to the slowly decaying electrons from the Starfish explosion.   \cite[From][]{Brown:1966}
\label{fig:andre8}
}}
{\includegraphics[width=0.9\textwidth]{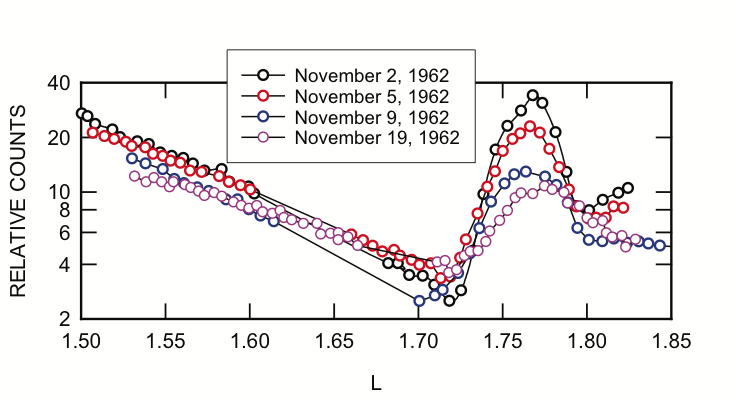}}
\vspace{-5mm}
\end{center}
\end{figure}

The third Soviet high-altitude explosion on November 1, 1962 created a narrow band of high energy electrons around L = 1.76 \cite[]{Brown:1966}.  \figurename~\ref{fig:andre8} shows the peak in this belt at four epochs after the explosion, at one, four eight and eighteen days after the blast, as a function of L.  As noted by \cite{Brown:1966}, the position of the peak in L remains constant while the intensity decreases by about a factor four.  At the same time, the full-width half-maximum of the peak increases monotonically from 0.037 to 0.052, implying the diffusion across L-shells of the electrons in the belt.  While this is a good illustration of the process well observed in the natural radiation belts \cite[e.g.][]{Williams:1968}, it does not completely explain the reduction in flux in the Soviet artificial radiation belt, as the broadening of the peak accounts for only 30\% of the losses.

\begin{figure}[t]
\floatbox[{\capbeside\thisfloatsetup{capbesideposition={left,top},
capbesidewidth=0.3\columnwidth}}]{figure}[\FBwidth]
{\caption{
A schematic representation of the evolution of the Starfish radiation belt over a year.  The decay continued for more than two years, until observations became difficult \cite[from][]{VanAllen:1964}. (The quantity plotted is according to the original labelling by James Van Allen.)
\label{fig:andre9}
}}
{\includegraphics[width=0.65\textwidth]{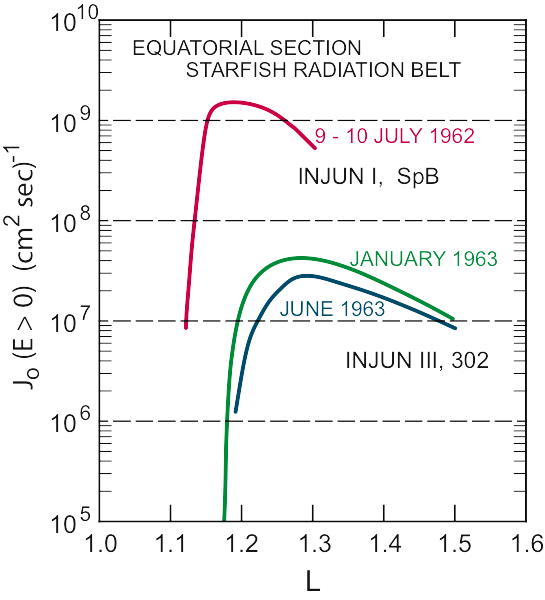}}
\end{figure}

\begin{figure}[h]
\floatbox[{\capbeside\thisfloatsetup{capbesideposition={left,top},
capbesidewidth=0.3\columnwidth}}]{figure}[\FBwidth]
{\caption{
Long term survival of the artificial radiation belt created by the Starfish detonation, observed by the energetic electron detector on the Relay II satellite.  The clustering of the data points is due to the precession of the satellite orbit.  The solid lines are least square fits of exponentials to the data.  \cite[From][]{Brown:1966}
\label{fig:andre10}
}}
{\includegraphics[width=0.65\textwidth]{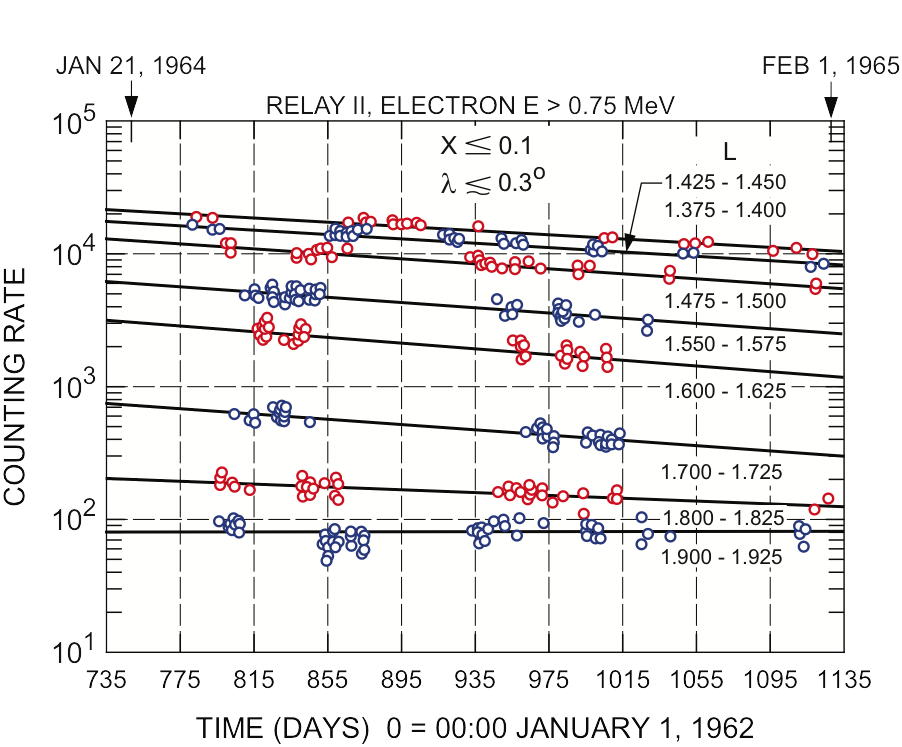}}
\end{figure}

A summary sketch of the Injun I and Injun III observation is shown in \figurename~\ref{fig:andre9} \cite[from][]{VanAllen:1964}.  The peak fluxes decreased by close to two orders of magnitude, by the erosion of the artificial belt at the low L values where atmospheric losses operated.  Removing that part of the artificial belt left the more stable belt at about L = 1.3 to persist for at least a year.  The complementary longer term observations by the Relay II satellite, launched on January 21, 1964 and fully equipped to measure the trapped radiation belts particles, including the remnants of the artificial belt populations, are shown in \figurename~\ref{fig:andre10} \cite[from][]{Brown:1966}.  The interesting aspect of these observations is that where the earlier decay observations showed decay constants of months at most, the still decreasing fluxes through to the end of 1964, more than two years after Starfish, now show that the time constants are in fact longer, as much as two years.  It is clear that the artificial belts created by the high altitude nuclear explosions endured at detectable levels for several years in a range of L-values about 1.5 to 1.8.  Both below and above these values loss mechanisms actively removed the excess population.  In the more stable range, however, it is likely that cross-L diffusion slowly moved the particles towards regions where more effective loss mechanisms finally removed them.

Finally, it is instructive to compare the exceedingly high energetic electron fluxes injected by the high-altitude nuclear explosions to today's state of the locations of the erstwhile artificial belts.

\begin{figure}[b]
\begin{center}
\includegraphics[width=0.9\columnwidth]{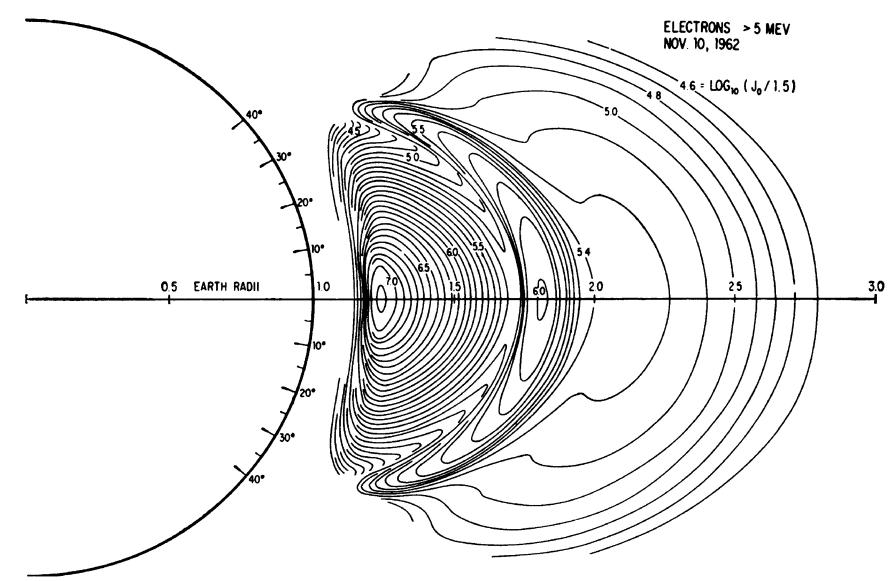}
\caption{Spatial distribution of high-energy electrons on 10 November 1962. The intensities represented by adjacent contours differ by a factor of 1.259 \cite[From][]{McIlwain:1963}.
\label{fig:baker2}
}
\end{center}
\end{figure}

Figure~\ref{fig:baker2} \cite[taken from][]{McIlwain:1963} shows that in mid-November 1962 (i.e., following the Starfish and major Soviet exoatmospheric tests), the omnidirectional flux of electrons with E$>$5 MeV exceeded $10^7$ cm$^{-2}$ s$^{-1}$ at L$\sim$1.2. The study by \cite[]{Brown:1966} similarly showed (see his Fig. 9) that electrons with E$>$1.9 MeV were above $10^7$ cm$^{-2}$ s$^{-1}$ for 1.2$\lesssim$L$\lesssim$2.5 and had a peak intensity of $>$$10^8$ cm$^{-2}$ s$^{-1}$ at L$\sim$1.4 in late October 1962.

\begin{figure}[h]
\begin{center}
\includegraphics[width=0.9\columnwidth]{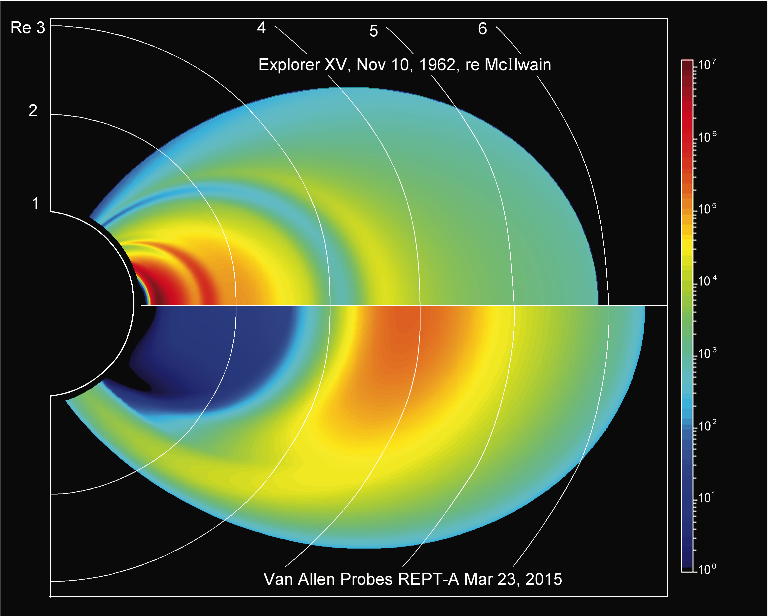}
\caption{
Comparison of Explorer XV data from 10 November 1962 with Van Allen Probes data taken with the Relativistic Electron-Proton Telescope (REPT) on 23 March 2015. The Explorer XV data are adapted from  \cite{McIlwain:1963} (see \figurename~\ref{fig:baker2}) and the Van Allen Probes data are adapted from measurements described in \cite{Baker:2016}. It is noted that the March 2015 outer zone fluxes of E$>$5 MeV electrons were more intense and broader in spatial extent than were such electrons in the November 1962 case. On the other hand, the inner zone electron fluxes in the 1962 post-Starfish era were at least 10 million times more intense than fluxes seen in the present-day magnetosphere. Such high fluxes were deadly to orbiting spacecraft of the early 1960s (as described in the text).
\label{fig:baker1}
}
\end{center}
\end{figure}

\figurename~\ref{fig:baker1} compares the post Starfish observations with the modern equivalent measurements from the NASA Van Allen Probes mission for a period (March 20--23, 2015) just following the most powerful geomagnetic storm of the last decade  \cite[see][]{Baker:2016}. The Van Allen Probes data show quite graphically that the ``natural'' radiation environment at present has no significant fluxes of E$>$5 MeV electrons at L$\lesssim$2.5 \cite[see, also][]{Fennell:2015, Li:2015}. The inner zone electron fluxes in the 1962 post-Starfish era were at least $10^7$ times more intense than even the highest fluxes seen in the present-day magnetosphere.

\section{Damage to satellites}
\label{sec:satellites}

Satellites in orbit at the time of the nuclear tests experienced energetic particle fluxes that were intense not only in the short term, but -- as can be seen in \figurename~\ref{fig:andre6} -- remained significant long enough for the accumulated radiation dose to cause damage.  Starfish Prime, having been the largest explosion, with the best documented consequences in terms of the artificial radiation belt created is also the one that had the best documented consequences in terms of satellites destroyed or damaged.  A first comprehensive review of the effects on satellites was given by \cite{Hess:1963}.  The near-immediate satellite losses due to Starfish: Ariel-1, TRAAC and Transit-4B, had been documented earlier \cite[]{Durney:1962, Hess:1962, Fischell:1962b}. 

\begin{table*}[!]
\begin{center}
\caption{Satellites in orbit and damaged by the high-altitude nuclear explosions}
\label{tab:damage}       
{\renewcommand{\tabcolsep}{3pt}
\resizebox{\columnwidth}{!}{%
\begin{tabular}{L{0.6in} R{0.4in} R{0.4in} L{1.6in} L{1.6in}}
\hline\noalign{\smallskip}
Satellite & Launch & End & Damage history & Comments \\
\noalign{\smallskip}\hline\noalign{\smallskip}
Injun I & Jun 29, 1961 & Mar 6, 1963 
  & Likely damage to solar cells and transmitter electronics, but not documented as associated with the Starfish belt.
  & Low power drain probably limited the damage due to reduced power from the solar cells \\
\noalign{\medskip}
TRAAC & Nov 15, 1961 & Aug 12, 1962 
  & Failed due to radiation damage to solar cells 
  & Solar cell test circuits enabled monitoring and diagnosing the failure of the power system \\
\noalign{\medskip}
Transit-4B & Nov 15, 1961 & Aug 2, 1962 
    & Failed due to radiation damage to solar cells 
    & Solar cell test circuits enabled monitoring and diagnosing the failure of the power system \\
\noalign{\medskip}
OSO-1 & Mar 7, 1962 & May 1964 
  & Eventual failure was due to solar cell damage 
  & Operated only intermittently after the tape recorder failure in May 1962, prior to Swordfish. \\
\noalign{\medskip}
Ariel-1 & Apr 26, 1962 & Aug 5, 1962 
  & Failed due to radiation damage to solar cells.  Command system electronics damaged; intermittent failures. \\
\noalign{\medskip}
Kosmos-5 & May 28,1962 & Decay date: May 2, 1963 & Last reported data in the literature is from November 1962.  Reportedly damaged by Starfish.\\
\noalign{\medskip}
Telstar-1 & Jul 10, 1962 & Feb 21, 1963 
  & Well documented damage to the solar cells, but eventual failure was caused by radiation damage 
     to transistors and other electronic devices 
  & This spacecraft was extensively instrumented to measure radiation damage, both to solar cells and to active devices 
  (transistors) in the electronics. \\
\noalign{\medskip}
Alouette-1 & Sep 29, 1962 & Jan 28, 1968 
  &Reportedly no adverse effects from the Starfish radiation 
  & Very conservative power supply design, allowing for 40\% degradation of solar cell performance \\
\noalign{\medskip}
Explorer XIV & Oct 2, 1962 & Aug 11, 1963 
  & Intermittent electronic failures in the spacecraft encoder \\
\noalign{\medskip}
Explorer XV & Oct 27, 1962 & Jan 30, 1963 
  &  Intermittent electronic failures in the spacecraft encoder, failure due to under-voltage caused by radiation damage \\
\noalign{\medskip}
ANNA-1B & Oct 31, 1962 & ? 
  & Deterioration of the solar cells noted, but not fatal 
  & Flew the first GaAs solar cells in space, making the solar array power source more resistant to radiation \\
\noalign{\smallskip}\hline
\end{tabular}
}}
\end{center}
\end{table*}

\cite{Hess:1963} lists 6 satellites that sustained damage from the energetic particles generated by Starfish.  These are Ariel-1 TRAAC, Transit-4B, Tiros 5 and OSO-1.  More extended lists were published later by \cite{Wenaas:1978} and \cite{Conrad:2010}.  The most relevant contemporary missions that were affected (or not) by the increased radiation due to the artificial radiation belt are listed in Table~\ref{tab:damage}, together with their failure modes and any reason for mitigation for those missions that had their useful life shortened but did not immediately fail after Starfish.  

The nuclear explosion on July 9, 1962 at 09:00:09 GMT was very quickly sensed by the radiation detectors on all five spacecraft: Kosmos-5, Ariel-1, TRAAC, Transit B, and Injun I then operating. 

The two satellites closest to the explosion were Ariel-1 and Kosmos-5 \cite[]{Durney:1962, Galperin:1964}.  Kosmos-5 was on a meridian $\sim75^\circ$ to the west and Ariel-1 at $\sim27.5^\circ$ also to the west of Johnston Island. Their latitudes and height were $44.9^\circ$N and 1442 km (Kosmos-5) and $52^\circ$S and 819 km (Ariel-1).  Their distances to Johnston Island were very comparable: 7,400 km for Ariel-1 and 7,500 km for Kosmos-5.  While Ariel-1 was closer to the magnetic meridian of the detonation site, on shell L$\sim$5, the greater height of Kosmos-5 allowed viewing above the site over the horizon at a height of $\sim$1,200 km, while for Ariel-1 direct visibility over the site was above 2000 km.   These differences probably explain the differences in the observations as well as their interpretation, although both provide evidence of the geographically very broad reach of the immediate effects.

\begin{figure}[h]
{\caption{
High time resolution observations on the immediate radiation detected following the Starfish explosion.  (a)  Observations by Ariel-1, the first spacecraft to be fatally damaged by the Starfish explosion. In the first minutes after the explosion, both the Geiger counter and the {\v{C}}erenkov detector saturated, as Ariel crossed the magnetic meridian of the explosion site within two minutes.  \cite[]{Durney:1964}. (b) Observations by the Geiger counter on the Kosmos-5 satellite showing the near-instantaneous increase of three orders of magnitude or more.  The instrument may have entered a near-saturation mode, but its counting rate was corrected using the cross-calibration with the non-saturating fluorescent electron detector as shown by the red line.  \cite[]{Galperin:1964}.
\label{fig:sat1ab}
}}
{
\includegraphics[width=1\textwidth]{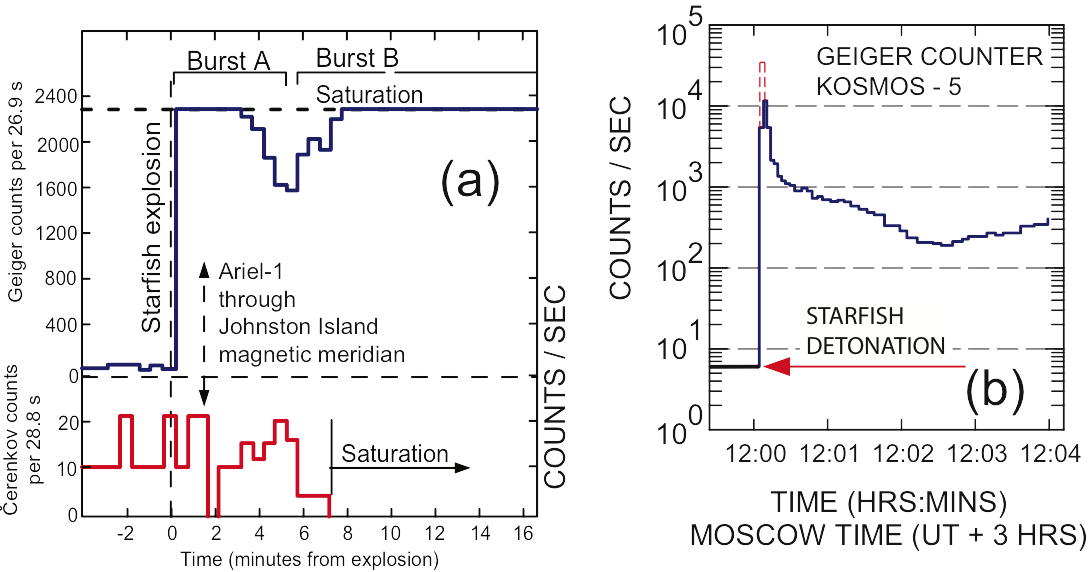}
}
\end{figure}

A plot showing the observations by Ariel-1 immediately following the detonations are shown in \figurename~\ref{fig:sat1ab}a.  Ariel-1 \cite[]{Baumann:1963} was a joint US/UK mission carrying, among other instruments, two energetic particle detectors with the original objective of measuring the fluxes of cosmic rays.  The characteristics of the instruments were described by \cite{Durney:1962} together with the main features of the observations.  The Geiger counter responded to both protons with energy E$>$43 MeV and to electrons with energy $>$4.7 MeV.  In \figurename~\ref{fig:sat1ab}a, it can be seen that the Geiger counter was saturated within 20$\pm$10 seconds of the explosion.  It was concluded at the time that this first burst of saturation was most likely to be caused by the magnetic disturbance propagating upwards of the explosion site at speeds $\sim$100 km/s that disturbed the outer radiation belt electrons and by lowering their mirror points effectively dumped these electrons at the location and height of Ariel-1.  This is supported by the proximity of the satellite to the magnetic meridian passing over the explosion site. The Geiger counter then saturated again, by then from the beta decay of the fission products of the bomb drifting eastward. 

The observations made by Kosmos-5 \cite[]{Galperin:1964} also make clear that there was an effectively immediate and very large radiation episode experienced far away from the Johnston Island site.  The energetic particle instrumentation carried on Kosmos-5 was described in detail by \cite{Galperin:1964}.  It consisted of a Geiger counter apparently very similar to those on the other, contemporaneous satellites, with a comparable shielding (3.4 g/cm$^2$ Pb and 0.8 g/cm$^2$ Al) and therefore had a comparable energy response to high energy electrons and protons.  The observations made immediately after the detonation are shown in \figurename~\ref{fig:sat1ab}b.  A first clear difference with the Ariel-1 observations is the apparent absence of saturation that allowed the initial peak intensity, about four orders of magnitude greater than the pre-starfish level, to be recorded.  Although, as explained by \cite{Galperin:1964}, the counter departed from the linear regime, it had been possible to calibrate the counting rate for high particle fluxes by the use of an additional instrument, an ``electron indicator,'' effectively an analogue electron detector using a fluorescent screen and a photomultiplier monitoring the input electron flux.  As the fluorescent detector did not saturate even in the maximum flux intensity regions of the artificial radiation belt as it evolved, the two detectors could be cross-calibrated. 

Even if the first explanation of the origin of the first burst of particles is the dumping of particles from the disturbance to the natural radiation belt, the sustained high intensities after the initial saturation are best explained by the decay of fission products.  This was the conclusion reached by \cite{OBrien:1962a}, on the basis of pre- and post-burst observations of the high energy particles on the Injun I satellite.  Their instrument, also a shielded Geiger counter, had a closely similar response (protons with E$>$45 MeV and electrons with E$>$6 MeV) to the Ariel-1 detector.  The first observations on Injun I were made at the first pass over a ground station in Southern Rhodesia (now Zimbabwe) at 09:42 to 09:50 GMT, about 45 minutes after the explosion.  By that time, electrons from fission products had time to form the eastward drifting shell of fission product electrons.

Another explanation of the very fast, first burst of saturation seen by Ariel-1 could be associated with the ionization and very fast motion of the fission products along the magnetic field lines, as reported later by \cite{DArcy:1965}, based on the interpretation of the high energy gamma radiation measurements near the magnetically conjugate point of the explosion site.  Their argument is that a large population of quickly trapped electrons was generated by the decay of the fission products during the flight along the magnetic field line.  The rise time of the excess gamma rays was very short, reaching a peak at $\sim$20 seconds from the explosion, showing that the decay-product electrons reached far from the explosion site to generate the saturation seen in the Ariel-1 detector.  These arguments also apply to the results reported by \cite{Galperin:1964} about the peak increase of radiation immediately following the detonation.

\cite{DArcy:1965} also suggest that the interaction of the plasma associated with the explosion with the geomagnetic field could have caused a mixing on neighboring flux surfaces, so that the original Ariel-1 explanation for the first burst of saturation, involving the disturbance in the magnetic field lines, was a factor in the minutes after the explosion.  But given the very intense gamma ray radiation described by \cite{DArcy:1965}, it is also possible that these contributed (maybe significantly) to the initial burst of saturation observed by Ariel-1 \cite[]{Durney:1964}.

\begin{figure}[h]
{\caption{
(a) The Geiger counting rate observed on the Ariel-1 satellite.  In the hours following the explosion the instrument regularly saturated on each orbit as the spacecraft crossed the forming artificial radiation belt.  (b) In the long term observations of the stable high energy cosmic rays by Ariel-1, the high energy electrons injected by the explosion stand out even at high value L-shells.  [Figures from \cite{Durney:1962} and \cite{Elliot:1966}.]
\label{fig:balogh2ab}
}}
{
\includegraphics[width=1\textwidth]{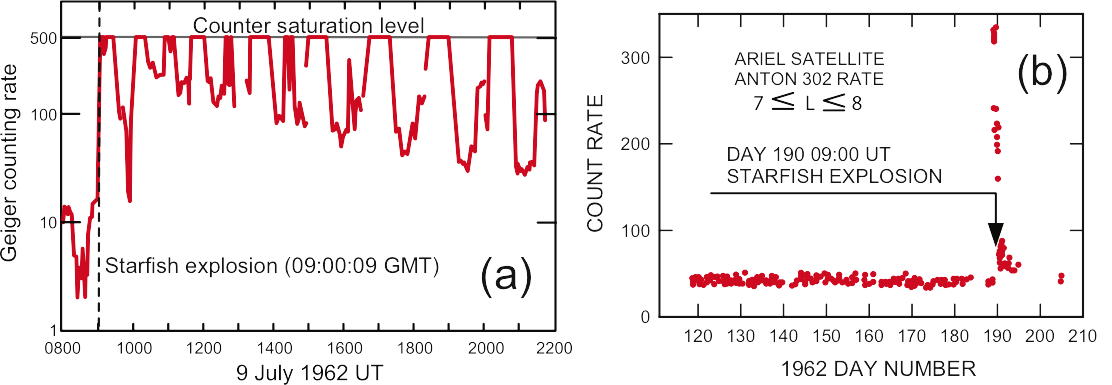}
}
\end{figure}

In the left hand panel (a) of \figurename~\ref{fig:balogh2ab}, the periodic passage through the then forming and evolving artificial electron belt, when the detector saturated, can clearly be seen.  The minima occur when the satellite moved to the outer radiation belt that had large fluxes of the transient energetic electrons from the explosion out to L = 4 and beyond.  However, the decrease in the intensity of the minima shows that at these high L-shells the fluxes decayed quite rapidly.  The right hand panel (b) shows the average counting rates at the highest L values (L $\sim$7 to 8), normally measuring the cosmic ray intensity.  The peak at the time of the Starfish explosion provides another proof that the high intensity radiation reached far from the explosion site, but that at these heights the radiation was only short-lived.

The complexity of the early phase of the artificially formed radiation belt is illustrated by the possibility that a short-lived second ``belt'' was formed \cite[]{Pieper:1963b} below the longer lasting artificial belt which, however, may not have survived the eastward drift of the trapping shells past the South-Atlantic anomaly.  This population of electrons had also been found by Ariel-1 \cite[]{Durney:1962} and Kosmos-5 \cite[]{Galperin:1964}.  Further analysis of the Ariel-1 observations \cite[]{Durney:1964} also found evidence for a short-lived (about one day) shell of energetic electrons, probably due to fission products remaining in the vicinity of Johnston Island after the explosion.

Thus the early radiation exposure of the satellites in orbit on July 9, 1962 may have been markedly different, depending on their actual orbits and the timing of their orbits with respect the explosion site and their magnetic connectivity.   This, together with their somewhat different design and components (in particular the solar cells) has led to the differences in the timing of their failures.  

All evidence points to Ariel-1 and Kosmos-5 being the most exposed, particularly early after the explosion.  Ariel-1 was the first spacecraft to show functional disturbances in the electronics of both the spacecraft system and in the scientific instruments.  Intermittent electronic failures were noted within a few days, but 104 hours after the explosion the power supply showed the first undervoltage condition \cite[]{Wenaas:1978}, triggering a protective switch off of the payload, resulting from a 25\% drop in power from the solar cells.  Although the satellite partially recovered, this condition dominated the next three weeks, finally leading to the end of the mission on August 5, 1962.  There is no concrete report on damage to Kosmos-5, although it is generally considered to have had its operational life shortened by the Starfish radiation exposure \cite[]{Wenaas:1978, Conrad:2010}.  It continued to provide data at least until late October 1962, as evidenced by the published observations.  It re-entered the atmosphere on 2 May 1963.

The TRAAC and Transit-4 satellites were launched simultaneously into the same orbit (1100 km $\times$ 950 km, 32$^\circ$ inclination) on November 15, 1961.  TRAAC was a satellite to test the concept of gravity gradient attitude control, but also carried energetic proton and electron detectors \cite[]{Fischell:1962a}. Transit-4-B was a US Navy navigation satellite, an early, experimental member of the long-running Transit series \cite[]{Danchik:1998}.  Transit-4B carried no scientific instruments and was powered by both solar cells and one of the first Radioactive Thermoelectric Generators, the SNAP-3.  The two satellites carried different circuits to measure the output of the solar cells.  This monitoring function showed that from launch to July 9, Transit-4B and TRAAC solar cells lost, respectively, 17\% and 18\% of their initial power output.  However, following the Starfish explosion, the deterioration in the Transit-4B solar cell output was 22\% in 20 days. The equivalent figure for TRAAC was 22\% in 28 days.  Transit-4B ceased transmitting on August 2, 1962, and TRAAC also stopped transmitting on August 14, 1962 \cite[]{Fischell:1962b}.  It should be noted that the nuclear power source on Transit-4b that could have continued to power the satellite, itself failed a month before the Starfish explosion, on June 12, 1962.

\begin{figure}[h]
\begin{center}
\includegraphics[width=0.49\columnwidth]{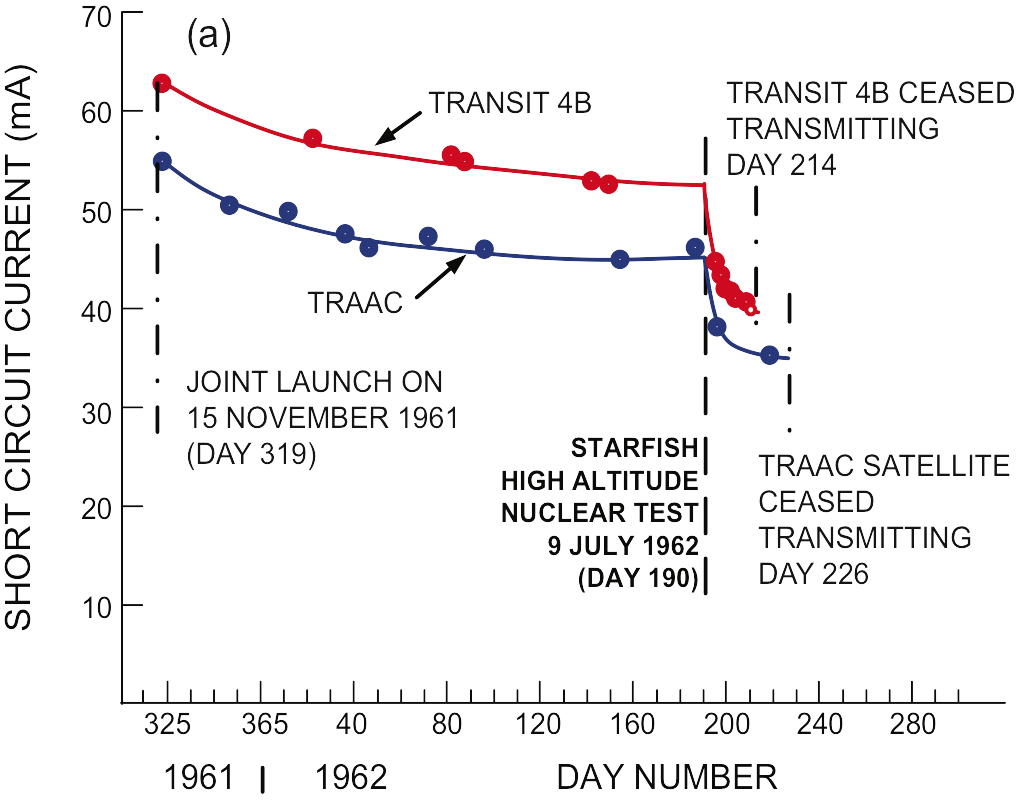}
\includegraphics[width=0.49\columnwidth]{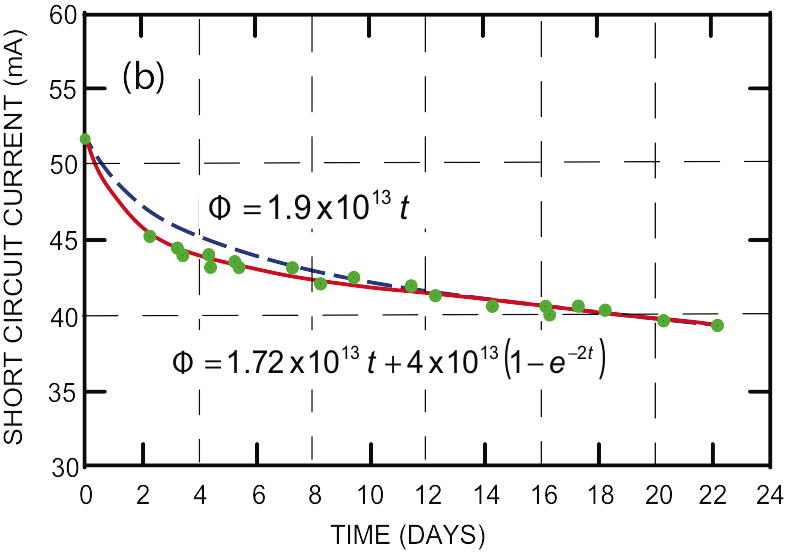}
\caption{Observation of the decreasing power from the solar cells on the TRAAC and Transit-4B satellites.  (a) Time history from launch, showing the gradual damage from the natural radiation environment, followed by the much accelerated loss of power after the Starfish explosion. [From Figure 5 in \cite{Fischell:1962b}.] (b) Transit-4B solar-panel performance after the Starfish event, showing that the performance loss is best modeled by an initial period of higher intensity radiation, followed by a constant radiation flux. [From Figure 7 in \cite{Fischell:1962b}.]
\label{fig:sat2ab}
}
\end{center}
\end{figure}

The time history of the output of the solar cells on TRAAC and Transit-4B is shown in \figurename~\ref{fig:sat2ab}(a) [Figure 5 in \cite{Fischell:1962b}].  The difference between the rates of deterioration in the natural environment and in the artificial radiation belt after Starfish is clearly shown.  In \figurename~\ref{fig:sat2ab}(b), the decrease in short circuit current (a measure of the power output) is shown in detail for TRAAC [from Figure 7 in \cite{Fischell:1962b}].  Two empirical curves were fitted to the observed measurements.  One (dashed line) assumed that there was a constant high energy electron flux incident on the solar panels, about $1.9 \times 10^{13}$ 1 MeV electrons cm$^{-2}$ per day.  This is to be compared with the average pre-Starfish value of $8.9 \times 10^{10}$ 1 MeV electrons cm$^{-2}$ per day, a factor 225 less than after Starfish.  The second empirical curve (solid line) assumes a constant flux of $1.72 \times 10^{13}$ 1 MeV electrons cm$^{-2}$ per day, with the addition of a flux component $4 \times 10^{13}$ 1 MeV electrons cm$^{-2}$ per day that decayed with a time constant of 12 hours.  This curve is a better fit, confirming that there was a large initial flux that decayed over the first two days.  It is this large initial component that was probably both space and time dependent that led, at least in part, to the different spacecraft being affected in different ways.  However, the very close history of TRAAC and Transit-4B is due to their proximity in orbit, as well to the similar components used in their construction.

The fourth spacecraft also in orbit at the time of Starfish was Injun I \cite[]{Pieper:1961}, extensively instrumented with radiation detectors.  Injun I was launched on 29 June 1961 into an orbit 1020 km $\times$ 820 km that was similar to Transit-4B and TRAAC, but with a higher, 67$^\circ$ inclination.  It was able to measure both the pre-Starfish \cite[]{OBrien:1962b} and post-Starfish energetic particle populations in the radiation belts \cite[]{OBrien:1962a, Pieper:1962, VanAllen:1963}.  The observations, possibly limited in spatial coverage by the orbit, identified the maximum intensities of the trapped electrons from fission products in a thin shell centered at L = 1.2, with omnidirectional intensities up to $10^9$ energetic electrons cm$^{-2}$ s$^{-1}$ shortly after the detonation on 9 July 1962.  This interpretation of the data was challenged by the Telstar-1 observations that showed a much broader artificial radiation belt out to L$>$2 \cite[]{Brown:1963a}. The high values of the energetic electron fluxes initially, of order $10^9$ electrons/cm$^2$/sec, were confirmed by \cite{Galperin:1964} who, however, also argue that these high fluxes had a greater spatial extent (up to L$\sim$2) than seen by Injun I which was presumably limited by its orbital coverage.  It should be noted that the Kosmos-5 Geiger counter was very similar in shielding and performance characteristics to the one onboard Injun I, so the differences in finding clearly represent the differences in orbital coverage, all the more so, that where they were at similar positions in orbit, the results are quite comparable.

Injun I operated until December 1962, and again, for a brief period, in early 1963.  However, \cite{VanAllen:1966} stated that ``useful data'' from Injun I was only received until late in August 1962.  The failure mode has not been reported, nor the design margin built into the power system.  However, the power used by Injun I was only 2 W \cite[]{Pieper:1961} and this very low power drain may have allowed the satellite to operate longer than TRAAC or Transit-4B, despite flying the same solar cells as the other two.

\begin{figure}[h]
{\caption{
The decay of excess radiation observed by the Kosmos-5 satellite following the Starfish explosion. (a) Excess electrons of energy $\leq$1 MeV measured by the electron indicator, an analogue, fluorescent detector near the magnetic equator, at L$\sim$1.2.   (b) The decay of the Geiger counter rate from the early, very high values at L$\sim$1.1, near the magnetic equator.  Note the log scale along the time axis. The initial measurements (red squares) can be in part attributed to gamma rays from fission products above the detonation site. [Both figures from \cite{Galperin:1964}.]
\label{fig:galperin2}
}}
{
\includegraphics[width=1\textwidth]{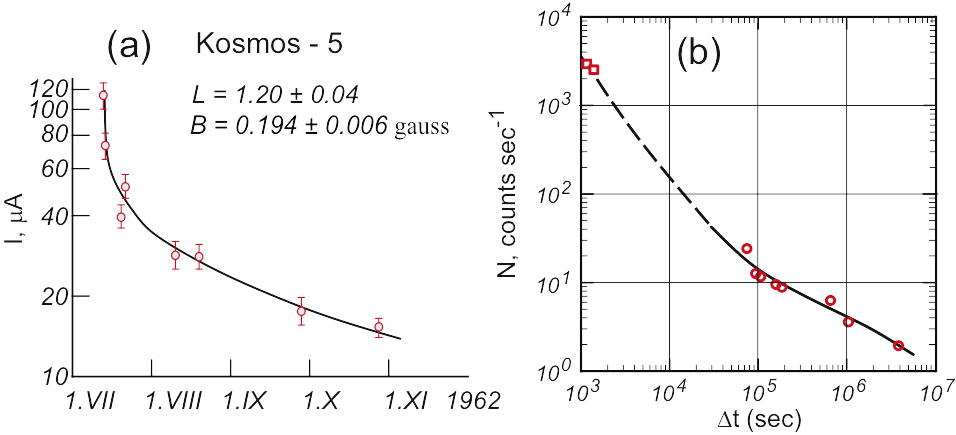}
}
\end{figure}

As Kosmos-5 continued to operate to at least October 1962, it was able to monitor the decay of the intensities observed with its instruments.  \figurename~\ref{fig:galperin2} shows the decrease in intensity by the two instruments onboard Kosmos-5 \cite[from][]{Galperin:1964}.  \figurename~\ref{fig:galperin2}a shows the decrease of ~ 1 MeV electron intensity measured by the ``electron indicator'' fluorescent detector.  \figurename~\ref{fig:galperin2}b is the decrease in intensity of the high energy electrons observed by the Geiger counter.  The very early highest intensity points are caused at least in part by delayed gamma ray fluxes from fission products.

Telstar-1 was launched on July 10, 1962, the day after the Starfish explosion.  It provided, thanks to its instrumentation and orbit, some of the most important and unique observations made in the aftermath of Starfish.  The satellite was considered an experiment in active satellite communications by the AT\&T company, who paid for the spacecraft, and paid NASA for its launch \cite[]{Dickieson:1963}. As outlined by Dickieson, the experiment had six objectives, the fifth of which was ``To gain a better numerical knowledge of the character and intensity of radiation in the Van Allen belt.''   Dickieson further wrote ``knowledge of the \dots radiation in the Van Allen belt was not adequate as a basis for design of a long-lived satellite.''  Hence, Telstar was the first commercial, non-scientific spacecraft that was instrumented to understand for operations and design what are now called effects of space weather on commercial technologies.

\begin{figure}[h]
\begin{center}
\includegraphics[width=0.53\columnwidth]{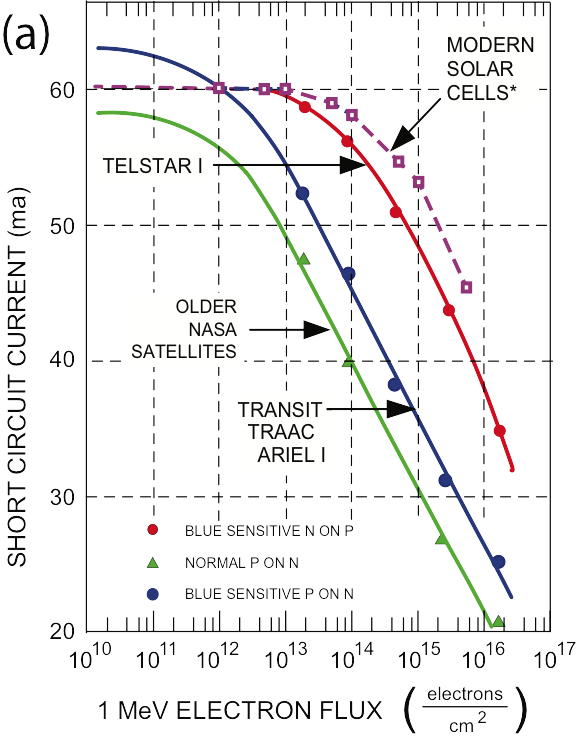}
\includegraphics[width=0.45\columnwidth]{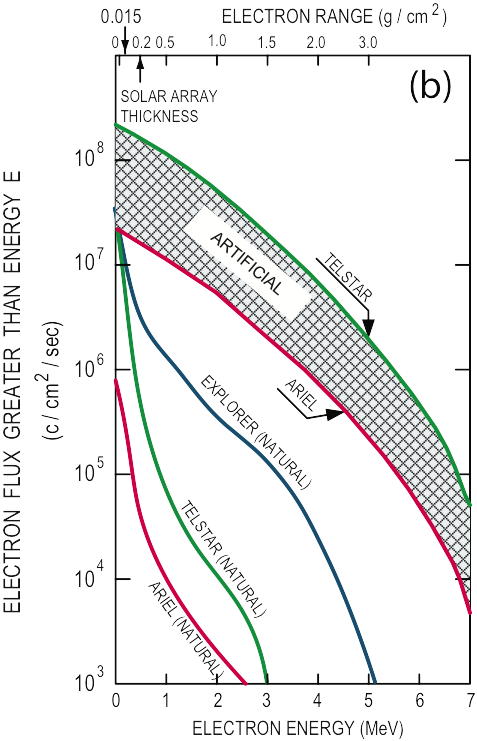}
\caption{(a) Solar cell damage curves. Based on Figure 3 in \cite{Rosenzweig:1963} and Figure 8 in \cite{Hess:1963}. The performance of the modern solar cells is taken from \cite{Xin:2014}. (b) Natural and artificial energetic electron fluxes showing cumulative spectra above 40 keV. From Fig. 4 in \cite{Wenaas:1978}.
\label{fig:sat3ab}
}
\end{center}
\end{figure}

Telstar was instrumented specifically to understand radiation effects on solar cells and on transistors, and to specifically measure the electron and proton radiation in the Telstar orbit \cite[]{Brown:1963a}.    The design and use of the four specific semiconductor particle detectors (three for protons, 2.4 to $>$50 MeV, and one for electrons, 250 -- 1000 keV) was considered a spin-off from the active nuclear physics program headed by Walter Brown at Bell Laboratories, and that had been initiated by Bell Labs management in the late 1950s [see Chapter 8 in \cite{Millman:1983}].   Semiconductor particle detectors had been invented at Bell Laboratories shortly after the invention of the transistor \cite[]{McKay:1949}, and the nuclear physics program extensively developed and employed them. 

The satellite's semiconductor detector array made crucial measurements of the fission-product electrons produced by the Starfish Prime high altitude nuclear explosion \cite[]{Brown:1963b}, see Section \ref{sec:bombs}.  In addition, Telstar-1 and the team behind it were extremely well equipped to assess the space radiation effects on solar cells and electronic devices (primarily transistors).  Solar cells had been developed in the first instance by Bell Labs scientists \cite[]{Chapin:1954}, so the solar cells used by Telstar-1 used new technology that also helped the radiation resistance, as is illustrated in \figurename~\ref{fig:sat3ab}(a) that compares the performance of solar cells used by contemporary satellites [based on Figure 3 in \cite{Rosenzweig:1963}].  \figurename~\ref{fig:sat3ab}(b) summarizes the energetic electron fluxes as a function of energy; the figure shows clearly the significantly higher flux rates that were experienced by Telstar-1 when compared, for instance, to Ariel-1.

\begin{figure}[h]
\floatbox[{\capbeside\thisfloatsetup{capbesideposition={left,top},
capbesidewidth=0.4\columnwidth}}]{figure}[\FBwidth]
{\caption{
The evolution of the average output of the current supplied by the solar cells on Telstar from launch to February 1963, corrected for mean solar distance.  The fit corresponds to the damage calculated for a mission-averaged 1 MeV equivalent energetic electron flux rate of $6\times10^{12}$ electrons cm$^{-2}$ per day.  The fit has been extrapolated to two years in orbit at the same energetic particle flux rate; the power output would then be reduced to 68\% of its initial performance.  [Taken from Figure 28, \cite{Brown:1963b}.]
\label{fig:sat4}
}}
{\includegraphics[width=0.55\textwidth]{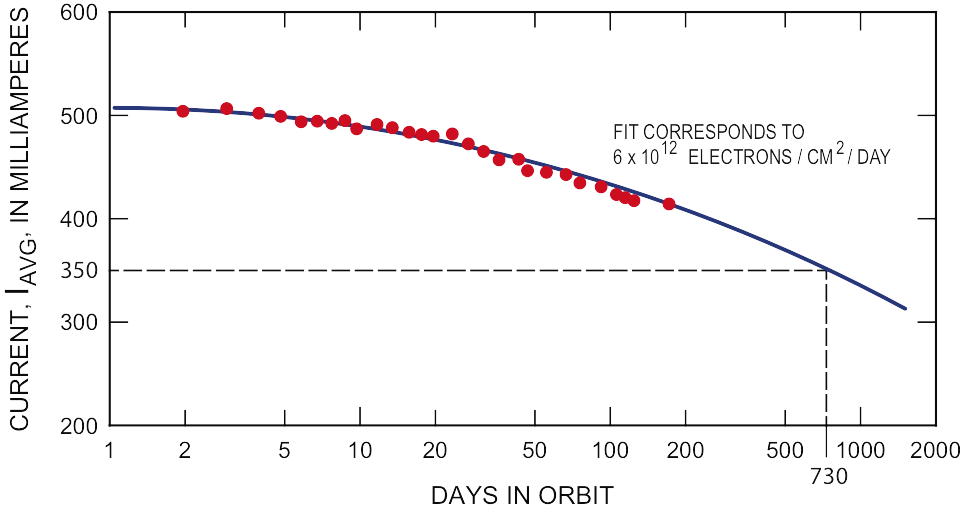}}
\end{figure}

The solar cell performance monitoring equipment on Telstar-1 logged the performance of the solar array, as shown in \figurename~\ref{fig:sat4} \cite[]{Brown:1963b}.  The rate of deterioration of the performance cannot be compared between pre- and post-Starfish epochs for Telstar-1, as was the case for TRAAC and Transit-4B as illustrated in \figurename~\ref{fig:sat3ab}.  However, the average energetic electron flux, referenced to 1 MeV electrons, $6 \times 10^{12}$ electrons cm$^{-2}$ per day is very high for the approximately 200 day lifetime of the mission.  This can be compared to the equivalent fit for the degradation of the TRAAC solar cells (see above) given as $1.9 \times 10^{13}$ 1 MeV electrons cm$^{-2}$ per day in the days following the Starfish detonation.

\begin{figure}[h]
\floatbox[{
\capbeside\thisfloatsetup{capbesideposition={left,top},
capbesidewidth=0.4\textwidth}}]{figure}[\FBwidth]
{\caption{
(a) Deterioration of transistor performance measured on Telstar-1.  These were specially instrumented ``damage-transistors'' for the purpose of observing the nature and magnitude of radiation damage. [From Figure 30, \cite{Brown:1963b}.] (b) Comparison of the radiation damage to transistors and solar cells.  Solar cells are more rugged, but spacecraft are linearly dependent on the power supplied by the cells, whereas circuits can be designed to be radiation resistant by taking into account the degradation in transistor parameters.  For today's spacecraft, solar cells remain critical, but the design and fabrication of electronic components have developed very significant radiation resistance.  [From \cite{Brown:1963a}.]
\label{fig:sat5ab}
}}
{
\includegraphics[width=0.55\textwidth]{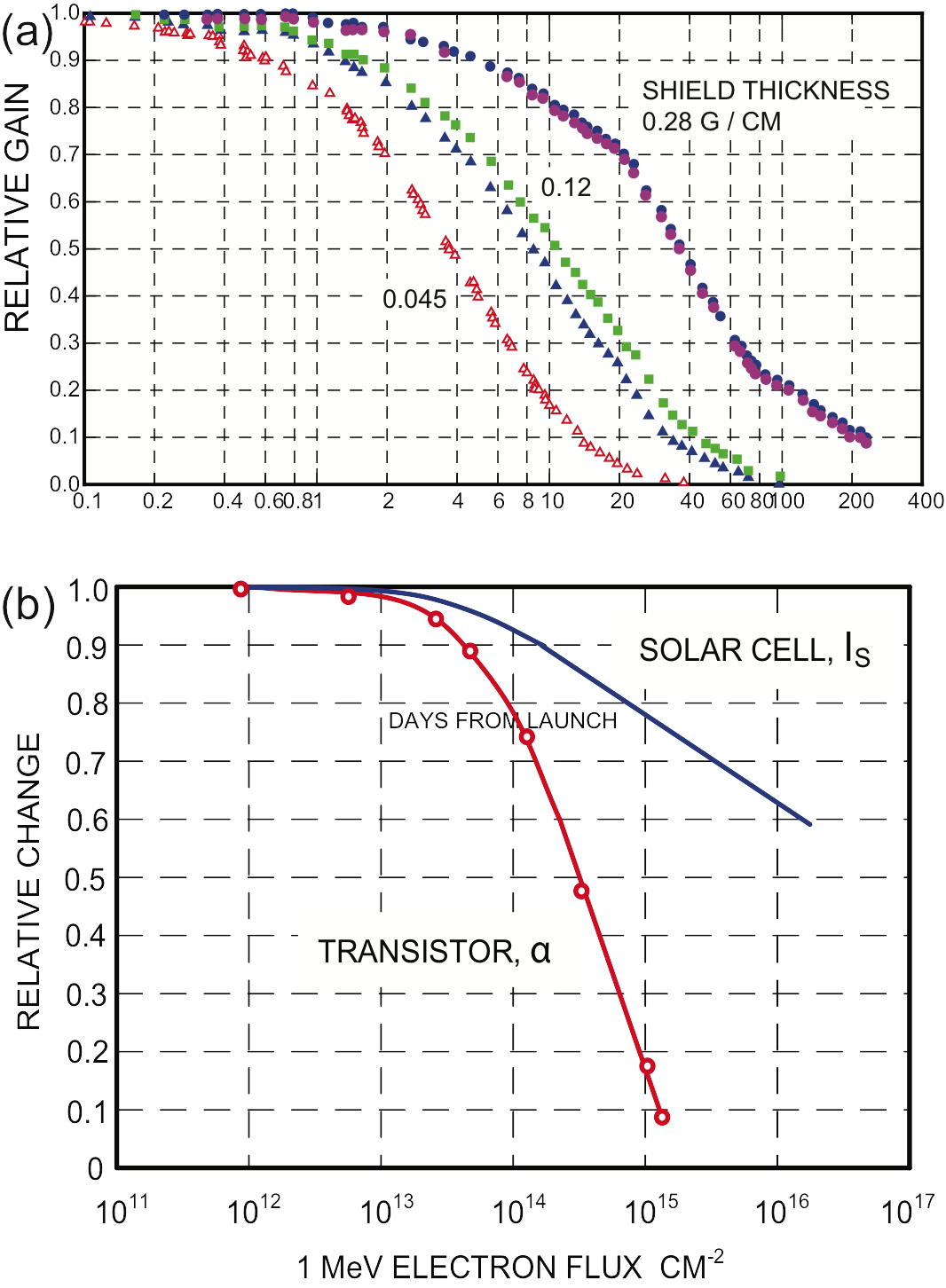}
}
\end{figure}

Despite the damage to the solar cells by the greatly increased radiation levels after Starfish, Telstar-1 in fact failed because of failures of electronic components in its command system.  Early indications were noted already a month after launch, but unexplained but fortunate partial recovery occurred so the mission could continue for several months more.  There was very little known about radiation damage effects in the active electronic components, diodes and transistors, of the time.  It was a far-sighted decision to include in the Telstar experiment some specific controlled test equipment to monitor the damage to transistors.  The seven so-called ``damage-transistors'' were mounted in a carefully controlled circuit to measure their parameters and with three different thicknesses of shielding \cite[]{Brown:1963a}.  The results are shown in \figurename~\ref{fig:sat5ab}(a), indicating significant damage to critical functional parameters of the transistors used.  In fact, a comparison between the damage susceptibility of transistors and solar cells at that time -- as illustrated in \figurename~\ref{fig:sat5ab}(b) taken from \cite{Brown:1963a} -- clearly shows the greater sensitivity of active electronic devices.  The fact that the spacecraft that failed did so because of the damage to their solar cells can be explained that most functional electronic components were usually well shielded within the spacecraft, even if this was not in the end the case for Telstar-1.

The radiation data from the Telstar sensors, together with extensive laboratory investigations of electron radiation effects with exposures as high as those measured in-situ, identified the malfunction and ultimate total failure of the satellite's command system in February 1963, some six months after launch \cite[]{Mayo:1963}.  Although the investigations pointed to surface damage in some of the transistors used from the enhanced radiation in the Van Allen belts, a highly ingenious procedure (foreshadowing many such efforts in later missions) had been worked out, using the detailed functionality of the command system that allowed circumventing to some extent the failure mode.  However, this recovery procedure also proved ineffective after repeated use.  It remains unclear whether the radiation damage process was identified with complete certainty [see doubts expressed by \cite{Wenaas:1978}], but as the design and fabrication of semiconductor components was evolving very fast even then, new devices came later to replace the transistors used in Telstar.

It is notable that early 1960s-vintage spacecraft were rather robustly built using the technology of the time. However, hazards in space operations and the effects of the space environment were only beginning to be quantitatively assessed at the time of Starfish.  The fact that several operational scientific, DoD and commercial spacecraft of that era failed due to damage from the artificial radiation belts is very sobering. Were nuclear explosions of the Starfish or Soviet high-altitude sort to be carried out today, the number of spacecraft that could be affected would be staggering. Given the much greater complexity of modern spacecraft electronics and the susceptibility of many such spacecraft to radiation dose effects \cite[]{Baker:2002}, the impact on modern infrastructure might be almost incalculable.


\section{Geomagnetic and Geophysical Signatures of High-Altitude Nuclear Explosions}
\label{sec:geomag}

Although the longest-lasting effect of the high-altitude nuclear blasts between 1958 and 1962 was the artificial radiation belt created by Starfish (and to a much lesser extent some of the others), the largest of them did generate other geophysical effects, including ionospheric and geomagnetic effects.  These were remarkable phenomena by both their propagation speed and (for the largest) their global reach.  However, in all cases these remained transients, lasting from minutes to hours, but nevertheless of sufficient magnitude that they qualify as ``anthropogenic'' space weather events.

The series of high-altitude nuclear explosions in 1958 (see Table \ref{tab:tests}) led to a series of studies based on the observation on magnetic and geophysical observatories where the detonations generated sudden, unusual signatures that were quickly associated with the nuclear tests.  The first account of such signatures was given by \cite{Cullington:1958} who described a highly unusual auroral display on August 1, 1962 at the tropical Apia Observatory, Western Samoa (latitude $~$14$^\circ$ South), together with geomagnetic signatures characterized as a Storm Sudden Commencement (SSC) at the same location.

\begin{figure}[h]
\floatbox[{
\capbeside\thisfloatsetup{capbesideposition={left,top},
capbesidewidth=0.3\textwidth}}]{figure}[\FBwidth]
{\caption{
The first geomagnetic signature of a high-altitude nuclear detonation.  Observations by the Apia La Cour magnetic observatory, Western Samoa, on August 1, 1958, following the TEAK nuclear test. The event was likened to a magnetic storm, preceded by a Storm Sudden Commencement \cite[from][]{Lawrie:1961}.
\label{fig:geomag1}
}}
{
\includegraphics[width=0.675\textwidth]{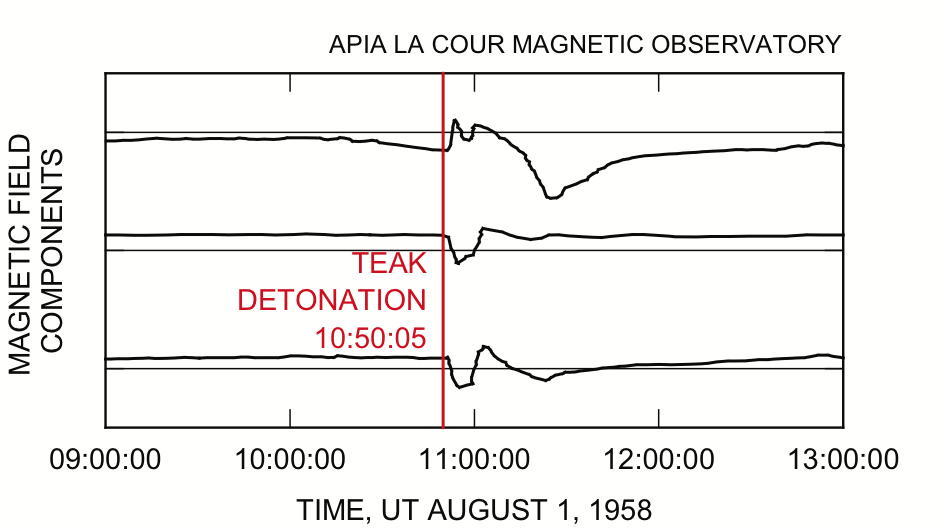}
}
\end{figure}

\begin{figure}[h]
\begin{center}
\includegraphics[width=0.8\columnwidth]{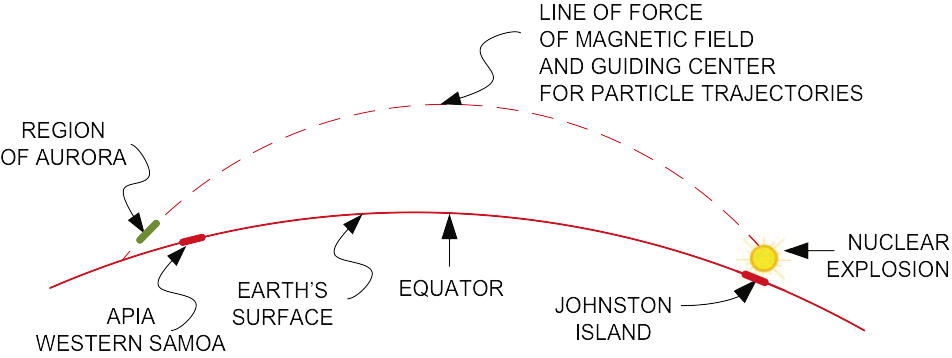}
\caption{Sketch of the origin of the auroral display on August 1, Western Samoa \cite[from][]{Kellogg:1959}.  Electrons that originated in the decay of fission products above the nuclear detonation site are guided by the Earth's magnetic field lines to the region magnetically conjugate where they cause an aurora when interacting with the atmosphere.
\label{fig:geomag2}
}
\end{center}
\end{figure}

The magnetic observations are shown in \figurename~\ref{fig:geomag1} \cite[from][]{Lawrie:1961}.  The Western Samoan aurora was explained by \cite{Kellogg:1959} and \cite{Elliot:1959} as having been caused by the nuclear explosion on August 1, 1958 over Johnston Island (Teak, see Table \ref{tab:tests}).  The aurora was all the more remarkable, because of the tropical location of the Apia Observatory.  \cite{Kellogg:1959} argued that the aurora was caused by energetic electrons traveling along the magnetic field lines from above the explosion site to its magnetically conjugate region, in this case Western Samoa.  \cite{Elliot:1959} recalculated the expected magnetically conjugate location of Johnstone Island using a different, surface-field-based magnetic field line model to get a better agreement with the auroral observations.  The sketch, in \figurename~\ref{fig:geomag2}, reproduced from \cite{Kellogg:1959}, illustrates the path of the particles that cause the aurora.  Note that these calculations predate the B-L coordinate system devised by \cite{McIlwain:1961} and later refinements that have provided much greater accuracy in magnetic conjugacy calculations.

\begin{figure}[h]
\floatbox[{
\capbeside\thisfloatsetup{capbesideposition={left,top},
capbesidewidth=0.3\textwidth}}]{figure}[\FBwidth]
{\caption{
Magnetograms from six stations distributed in distance from Johnston Island following the Argus III nuclear detonation on September 6, 1958.  It shows two groups of signatures; the first very prompt, the second showing reasonably clear velocity dispersion. The fitted velocities are by \cite{Berthold:1960}; for the prompt signal, \cite{Caner:1964} argued that it was arguably synchronous worldwide, as was later found for the prompt signature from Starfish.
\label{fig:geomag3}
}}
{
\includegraphics[width=0.625\textwidth]{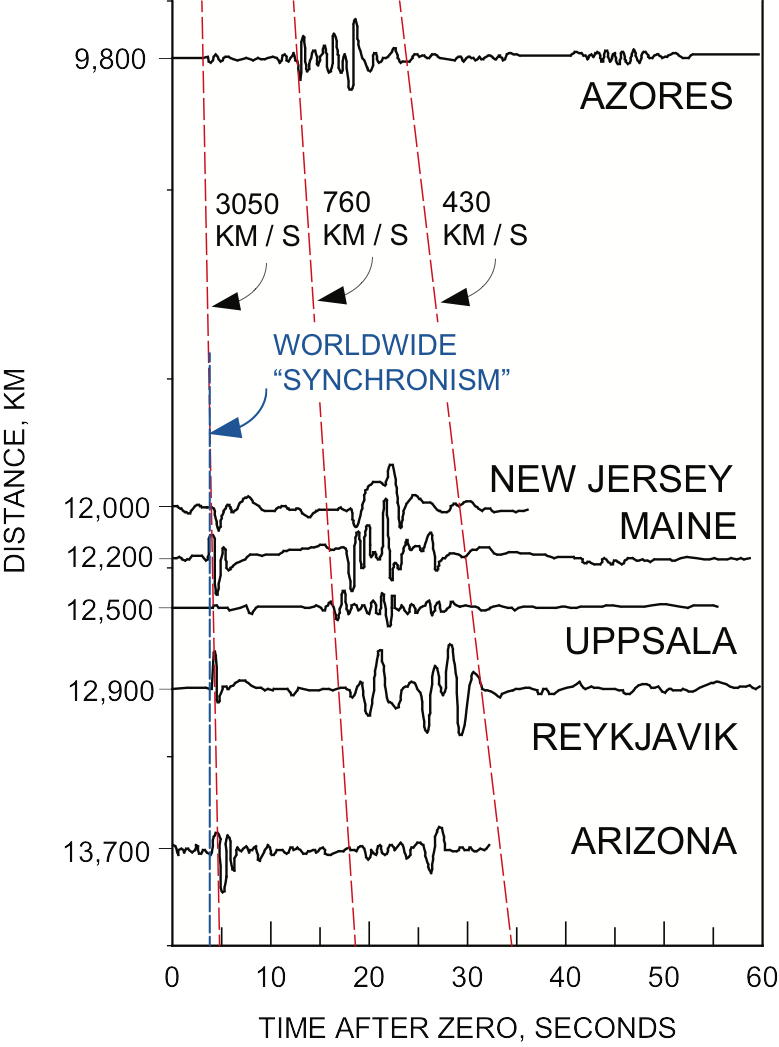}
}
\end{figure}

Geomagnetic signatures following the August 1, 1962 detonation were noted, away from the conjugate region in Honolulu, by \cite{Maeda:1959} who pointed out the similarities of the signatures (and their timing) with the observations in Western Samoa.  However, the world-wide scale of the geomagnetic signatures could only be observed after Argus III on September 6, 1958 and catalogued by \cite{Berthold:1960}.  They examined the geomagnetic signatures from six ground stations, well separated in longitude and covering a distance range of 9,800 to 13,700 km from the detonation site.  Plotting on the same scale allowed a comparison of the timing of the signatures and to deduce apparent velocities.  The results are shown in \figurename~\ref{fig:geomag3}.  Two signals were identified, a prompt one, with approximate velocity 3000 km/s and a delayed one, with velocity between 760 and 430 km/s.  Their conclusion was that these velocities are comparable with theoretical values calculated for hydromagnetic waves propagating in the ionosphere.  The propagation was approximately at constant speed; the magnitudes of the signals decreased with propagation direction away from the magnetic meridian of the detonation site.  The analysis for the onset was challenged by \cite{Caner:1964} who, in the light of the Starfish experience, proposed that the signal front was effectively synchronous worldwide.  That interpretation also fits with the (noisy) data -- the accuracy of the Azores data signal onset having been questioned by \cite{Berthold:1960}.  This is also illustrated in \figurename~\ref{fig:geomag3}.

These conclusions following the 1958 series of explosions came to be significantly revised after the analysis of the Starfish and other 1962 explosions.  The reason was the considerably increased height of Starfish, as well as its much greater yield than the Argus series, in addition to the greater readiness of ground-based instrumentation available for observations.  It is arguable that of these two parameters the height, at 440 km, was the more important.  While many of the associated geophysical phenomena were confirmed, the better coverage as well as the stronger signatures allowed more detailed conclusions to be drawn from the event.

\begin{figure}[h]
\begin{center}
\includegraphics[width=1\columnwidth]{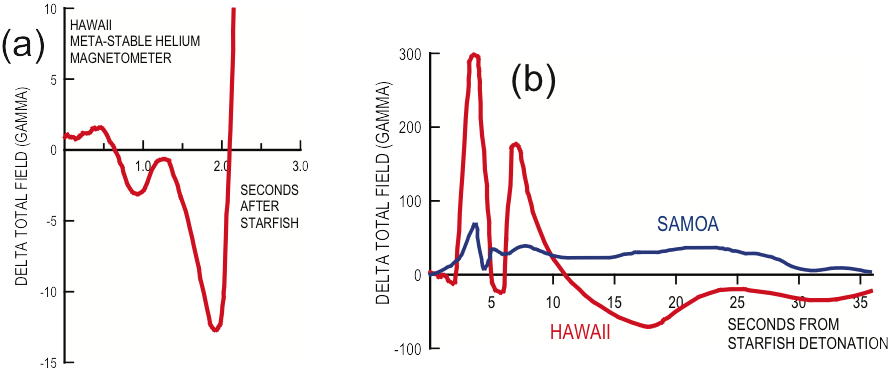}
\caption{The total magnetic field measured by meta-stable helium magnetometers following the Starfish detonation. (a) The initial response of a magnetometer located in Puako, Hawaii, showing an immediate response, less than one second after the detonation, followed by the second, high amplitude response after two seconds.  (b) The traces recorded in Puako, Hawaii and Tutuila, Samoa by identical meta-stable helium magnetometers.  Note that the initial responses at the two stations are in phase, although the Samoa peak is significantly smaller than that seen in Hawaii and the signals do not remain in phase after the initial, simultaneous peaks. \cite[From][]{Bomke:1964}
\label{fig:geomag4}
}
\end{center}
\end{figure}

\begin{figure}[h]
\begin{center}
\includegraphics[width=1\columnwidth]{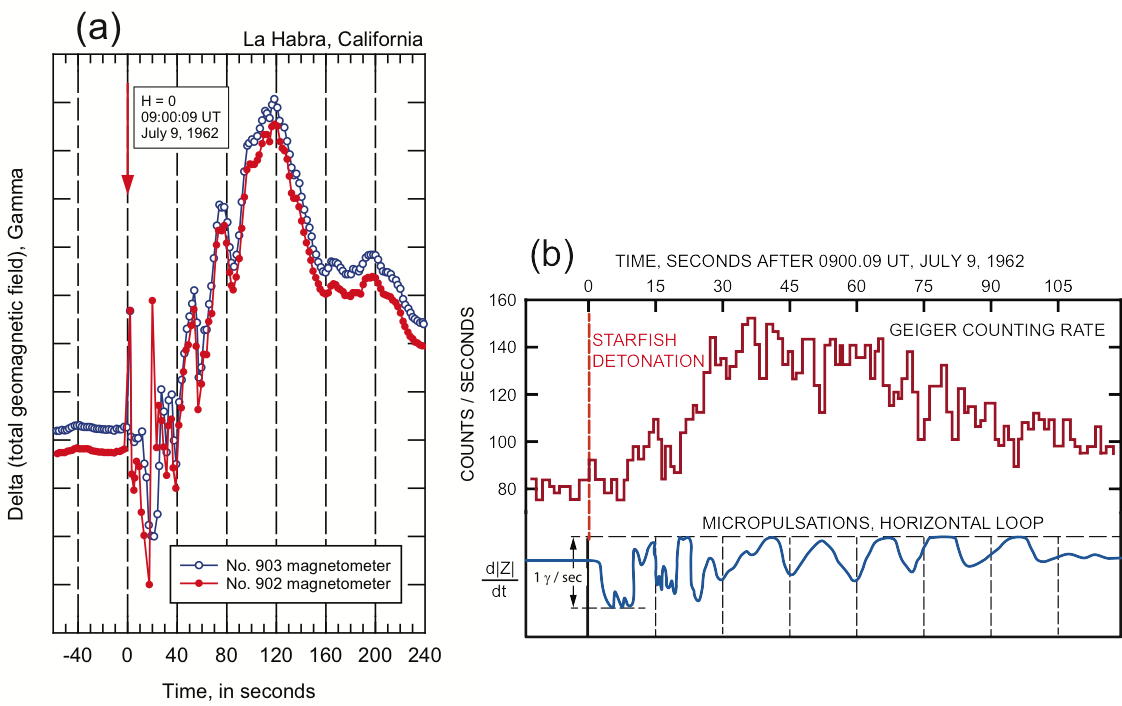}
\caption{(a) The magnetic disturbance observed with two rubidium-87 optically pumped vapour magnetometers at La Habra, California.  The onset of the disturbance is quasi-instantaneous and it lasts $\sim$7 minutes.  The rate of change at the onset was too fast for the sampling averages, hence the two instruments only track each other after about 20 seconds. \cite[From][]{Unterberger:1962} (b) The upper panel shows the counting rate of a balloon-borne Geiger counter at 80 g/cm2 atmospheric depth ($\sim$18,200 m height) near Hobart, Tasmania, following the Starfish explosion. In the lower panel, the record of micropulsations that followed the detonation is shown.  After the initial transient, irregular pulsation, a quasi-periodic oscillation was observed.  The 16-second period corresponds to the fundamental V-mode oscillation of the geomagnetic field line through Hobart. \cite[From][]{Edwards:1964} 
\label{fig:geomag5}
}
\end{center}
\end{figure}

A well-prepared and well-instrumented ground magnetometer network prepared by the U.S. Army Electronics Research and Development Laboratories was ready to record the geomagnetic signatures of the Starfish explosion on July 9, 1962.  As described by \cite{Bomke:1964}, all observatories made high-speed and detailed recording of the Starfish effects.  The onset of the disturbance at the Hawaii observatory is shown in  \figurename~\ref{fig:geomag4}a; it is seen to be effectively instantaneous on the time resolution of the magnetometer, better than 0.1 s.  The more intense signal arrives, with a very fast rise time, about 2 s after the detonation.  The full amplitude of the signal can be seen in \figurename~\ref{fig:geomag4}b; it peaked at about 4 s, but then, following another large-amplitude interval, decayed relatively fast on this scale.  The same figure also shows the corresponding signal recorded at the Western Samoa observatory.  The amplitude of the signal is significantly smaller, but the early time profile matches that observed in Hawaii: near-instantaneous onset, followed by a second, larger amplitude signal. The observations of the signature in the total field observed at La Hambra, California are shown in \figurename~\ref{fig:geomag5}a, illustrating again the very prompt onset, then the arrival of the higher magnitude signal and the concluding phase of the relatively short duration event.  The micropulsations observed in Tasmania, together with the simultaneously observed energetic electron counting rate onboard a stratospheric balloon are shown in \figurename~\ref{fig:geomag5}b.   

The geomagnetic disturbance caused by Starfish was reported to have been observed worldwide: in the Pacific and the continental United States \cite[]{Unterberger:1962, Breiner:1963, Bomke:1964, Miles:1964}, in Canada \cite[]{Baker:1962}, in Tasmania, Australia \cite[]{Edwards:1964}, in Peru \cite[]{Casaverde:1963}, in India \cite[]{Pisharoty:1962}, in France and in the Kerguelen Islands, South Indian Ocean \cite[both these by][]{Roquet:1963}.  A noted characteristic was the near-instantaneous arrival of the prompt onset, implying the propagation of the initial disturbance by electromagnetic means, at or near the speed of light.  \cite{Roquet:1963} report that at the two locations, in France and in the South Indian Ocean, the onset was timed at 09:00:08.8 $\pm$ 0.2 UT and 09:08:07.7 $\pm$ 0.5 UT, respectively, to be compared with the most accurate available timing of the Starfish detonation at 09:00:09 UT.

Following the global synchronism of the onset of the magnetic signal to within measurement accuracies, generally within a fraction of a second, a second, much higher magnitude signal was observed, as already indicated in connection with \figurename~\ref{fig:geomag4}a and -\ref{fig:geomag4}b.  The onset of this second signal was similarly global and synchronous and contained a significant oscillatory component with period about 3 to 4 seconds -- difficult to measure more precisely, due to the diversity of instrumental non-linearities that affected the observations.  The oscillations decayed rapidly, after five to eight periods.  

In reviews of the collected observations, \cite{Caner:1964} and \cite{Kahalas:1965} listed the characteristics of the two components of the prompt signal and offered explanations for them.  The first signal, at the time of the explosion, may be due to the absorption of the gamma rays in the atmosphere that generate an electromagnetic pulse which propagates in the earth-ionosphere cavity to generate a global signal.  The second component, detected globally after a $\sim$2 second delay, proved much more difficult to explain.  \cite{Caner:1964} considered all the alternatives and came to the conclusion that hydromagnetic standing waves along the magnetic field line from the explosion site are at the origin of waves that are converted to electromagnetic waves in the lower ionosphere.  These can then propagate in the ionosphere-Earth surface cavity at close to the speed of light.  The $\sim$2 second delay is then accounted for by the initial propagation delay along the magnetic field line.  

\begin{figure}[h]
\begin{center}
\includegraphics[width=1\columnwidth]{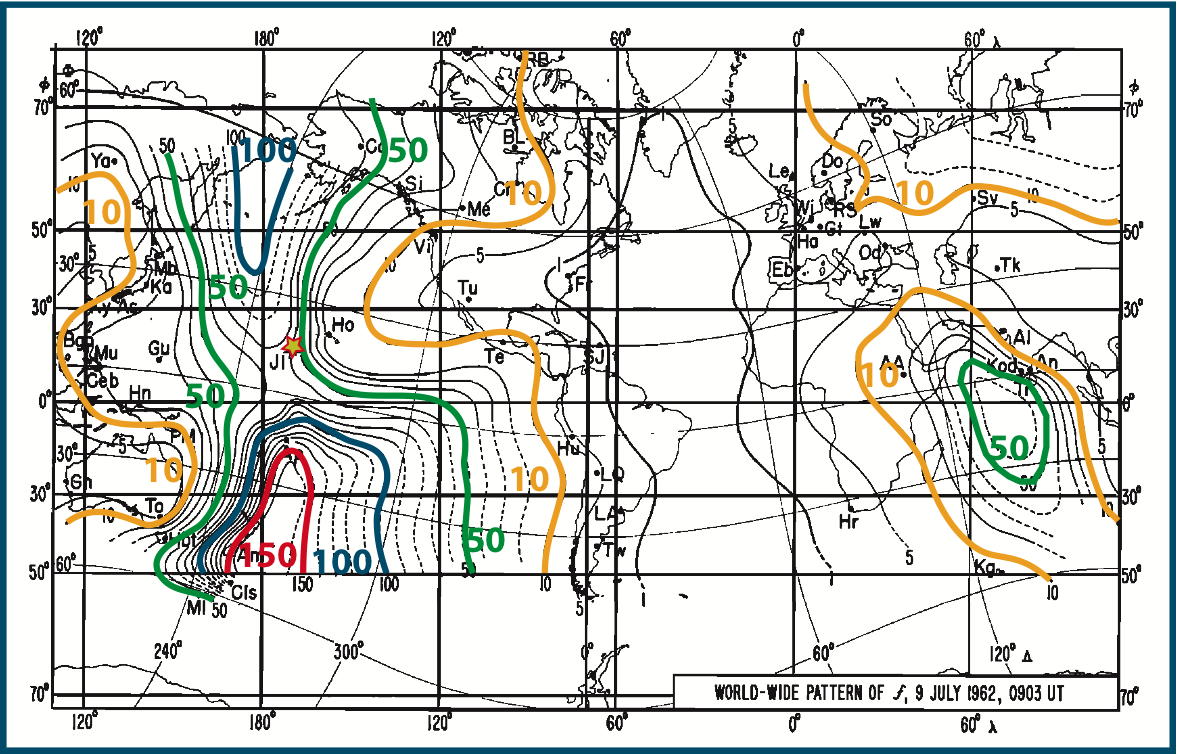}
\caption{Isopleths of  the amplitude  of  the total-disturbance magnetic vector  at  0903 UT  on July  9, 1962 (2  min 51 sec after  Starfish), plotted on a  Mercator  projection. The  detonation took place 400 km above Johnston Island (JI).  Values are in gammas. A contour interval of 10 gammas is used from the 10-gamma isopleth upward;  the color lines are 10, 50, 100, and 150 gammas \cite[from][]{Bomke:1966}.
\label{fig:bomke66}
}
\end{center}
\end{figure}

A very comprehensive modeling of the geomagnetic observations in the form of iso-intensity contours of the magnetic disturbances by \cite{Bomke:1966} have yielded the unexpected result that the maxima were in magnetically conjugate regions about the Johnston Island magnetic meridian plane, but at 45$^\circ$ magnetic latitude, rather than closer to the Johnston Island magnetic latitude of 14$^\circ$N (see \figurename~\ref{fig:bomke66}).  Their tentative explanation for this result is the coupling of magnetoacoustic waves with guided Alfv{\'e}n waves that is most effective at the conditions about 1000 km height, corresponding to the magnetic field lines at the higher latitude.  Auroral observations following Starfish \cite[]{Gabites:1962} support a very prompt activity at higher magnetic latitudes, in the mid-40$^\circ$ range, than the conjugate area of explosion site.  The conclusions of \cite{Bomke:1966}, together with the observation of auroral activity confirm that the magnetic field lines in the vicinity of the meridian plane of the explosion site were strongly disturbed to much greater heights and magnetic latitudes than was the case for the earlier, 1958 series of nuclear detonations.  In particular the sketch in \figurename~\ref{fig:geomag2} would need to be complemented by the inclusion of a significant volume of space reaching out to at least L = 5. This volume is now known to have corresponded to the diamagnetic cavity generated by the explosion, described below.  However, at the time, no information was available on the properties and dynamics of the diamagnetic cavity.

These considerations have received a late support, more than 40 years after the event, when \cite{Dyal:2006} published a detailed analysis of observations made at the time of the Starfish explosion by five fully instrumented sounding rockets near the Johnston Island site.  The data had not been in the public domain until this analysis was published and the observations throw an interesting light on the ``initial conditions'' following the explosion that could not be included in the contemporaneous studies in the 1960s.  The details relate to the dynamics and evolution of the diamagnetic cavity formed immediately after the explosion.  The cavity reached a length of 1840 km along the magnetic field lines and an extent in height of 680 km, reaching beyond 1000 km above the explosion site.  This extent of the cavity was reached in 1.2 seconds, and the bubble took about 16 seconds to collapse.  The intensity of electrons in the bubble was extremely high, of order $10^{13}$ cm$^{-2}$ s$^{-1}$ initially.  The magnetic field lines were clearly significantly distorted to relatively high L values above the explosion site, and relaxed once the cavity collapsed, but by that time high fluxes of electrons became trapped and formed the injection population for the artificial radiation belts as well as contributing to the generation of auroral precipitation in a broad range of magnetic latitudes, as described above.

In summary, the high-altitude nuclear detonations in 1958, but even more obviously those in 1962, and in particular the Starfish event on July 9, 1962, have generated widely-observed, worldwide geophysical signatures.  However, the geomagnetic effects were not long-lasting, unlike the artificial radiation belt.  The signatures were complex and despite the abundant contemporaneous literature, the physical phenomenology was at best tentatively identified.  It is not clear whether this was because the near-Earth space environment (including the ionosphere and exosphere) and its processes were not well understood at the time or because the observations, although plentiful, were neither sufficiently comprehensive nor sufficiently accurate.  In the absence of such events since then, it could be worthwhile to revisit the unique set of observations of the time to investigate if more could be learnt from them, particularly in the light of the observation made public by \cite{Dyal:2006}.


\section{Impact on the Power Grid}
\label{subsec:grid}

Since the advent of the electrical telegraph in the mid-19th century, it has been recognized that long metallic conductors grounded at both ends -- e.g., power grids, communication cables, pipelines -- can efficiently carry electrical currents that may exist in the Earth.  That is because the conductors fashioned by human technology have far less resistivity than does the Earth, and so the conductors provide an easy path for electrical currents flowing in the Earth.  Geomagnetic disturbances in Earth's magnetosphere produce variations in the current systems that flow in the ionosphere.   These ionosphere current variations produce in turn variations in the magnetic field at Earth's surface.    These magnetic variations generate electrical currents flowing in the solid Earth (illustrated in \figurename~\ref{fig:E3B} under a high altitude nuclear explosion). Significant electrical potential differences can occur over both short and long distances depending upon the scale sizes of the changing electrical currents in the ionosphere and the spatial conductivity of the underlying Earth, both horizontally and vertically.  

This geophysical effect was evident during the Starfish Prime event, when some 300 streetlights in Honolulu, about 1500 km from the detonation site, were extinguished. An analysis by \cite{Vittitoe:1989} indicated that the orientation of the magnetic fields from the bomb-produced EMP was consistent with the production of the streetlight outage \cite[]{Vittitoe:1989}.   The reported outage at the same time of a microwave telecommunications link between Kauai and other Hawaiian islands could have been caused by damage or outage of the electrical power to the microwave station by the EMP event. The damaging effects of EMP that can be expected on the electrical power grid of the nation (as well as on other infrastructures such as communications, banking and finance, transportation) are outlined in \cite{Foster:2004}.

While the streetlight failures in Oahu were apparently due to fuses in the system \cite[]{Vittitoe:1989}, the outages of electrical power systems due to geomagnetic activity can be produced by a number of different factors depending upon the grid system, the overlying ionosphere, and Earth's conductivity.    The Quebec outage of 1989 was caused by induced current-produced voltage depressions that could not be compensated by the system's automatic compensation equipment.  This event also saw generators tripping out of service and other effects in other systems \cite[]{Vittitoe:1989}.    The picture of the damage of a transformer at the Salem nuclear power plan in New Jersey in the 1989 event remains an iconic symbol of natural space weather effects on technical systems.   

The 2008 Workshop report from the National Academies \cite[]{Baker:2008} indicates that the least understood aspect of power system failure during a geomagnetic event is the permanent damage and loss of grid assets, especially large and costly transformers.   These transformers can experience excessive levels of internal heating that is caused by the induced Earth currents flowing through the devices.  The heating can produce melting and burn-through of copper windings.    Such damaged transformers cannot be repaired in the field, and have to be replaced.   The Report discusses the implications of the long lead times for producing new transformers, and therefore the implications of long outages if many are damaged in a geomagnetic event.  The same considerations apply to the societal implications of many damaged transformers that will occur in an EMP event, as outlined in \cite{Foster:2004}.  


\section{Other HEMP Impacts}
\label{subsec:myth}

Much of the literature on HEMP is either classified or not easily accessible. \cite{Savage:2010} and \cite{Foster:2008} discuss some ``real life'' effects. Here we briefly outline some of these phenomena.

\subsection{Cars}
\label{subsubsec:cars}

Some say that all vehicles traveling will come to a halt, with all modern vehicles damaged because of their use of modern electronics. A car does not have very long cabling to act as antennas, and there is some protection from metallic construction. As non-metallic materials are used more and more in the future to decrease weight and increase fuel efficiency, this advantage may disappear. More recently, the number of microprocessors in cars and the reliance on microprocessors in all motor vehicles has increased greatly.  Also, the sensitivity of the electronic circuitry to EMP has increased due to the use of smaller electronic components designed to operate on lower voltages. The fact is that not all cars need to be damaged to make the traffic stop. It is enough to permanently damage less than half (some sources say even 15\%) of cars to block cities, highways and supply lines.

\subsection{Mobile Devices}
\label{subsubsec:mobile}

About 95\% of these devices already have an internal electromagnetic interference shielding (which is there to protect the components from affecting each other). Plus, these devices are very small (compared to the power grid) so there is a very good chance that mobile devices can be used immediately after an EMP. But not to make calls or to search the internet. Telecommunication antennas will be unusable so there will be no signal of any kind. But if the phone is connected to the power grid -- charging -- at the moment when the EMP hits it will be damaged beyond repair.
 
\subsection{Pacemakers}
\label{subsubsec:pacemakers}

The hermetically sealed can is a good Faraday cage so the pacemaker will suffer no direct damage from an EMP. A bipolar system with, say, 5mm electrode separation will, with a major high-altitude EMP generating $\sim$25 kV/m in the northern US, deliver a voltage pulse to the circuitry of about 900V -- this will not damage a pacemaker, they are designed to cope with external defibrillator voltages. A unipolar system, where the electrodes may be 15cm apart, will generate a bigger voltage, maybe 4--5kV, to the generator, and this could be damaging.

\subsection{Airplanes}
\label{subsubsec:planes}

Older planes use hydraulics and cables attached to the pilot controls (with manual valve actuation and direct pressurization from the ram air turbine) which means they will still have semi-functional flight controls. Newer planes (almost all airlines) will be extremely difficult to control after all their electronic parts will be damaged. But even so, all Airbus and Boeing planes are demonstrated to be controllable with complete electrical failure. They would be extremely difficult to land, but it would still be possible. All planes will be turned into semi-controllable gliders... with newer planes having almost no control and older planes having almost full control over the plane. And pilots will face a very hard task: to land the plane, with highways being full of broken cars, no emergency services and if it is night then add no ground lights to this. But an EMP will not cause planes to enter in a spin and pancake into the ground.

\subsection{Maximum Conductor Length}
\label{subsubsec:conductor}

 There is a suggestion that equipment will not be damaged if all connected conductors are less than a specific length. Certainly shorter lengths are generally better, but there is no magic length value, with shorter always being better and longer not. Coupling is much too complex for such a blanket statement -- instead it should be ``the shorter the better, in general.'' (There can be exceptions, such as resonance effects, which depend on line lengths.)

\subsection{Turn Equipment Off}
\label{subsubsec:off}

There is truth to this recommendation (if there were a way to know that a burst was about to happen). Equipment is more vulnerable if it is operating, because some failure modes involving HEMP E1 phase trigger the system's energy to damage itself. However, damage can also happen, but not as easily, to systems that are turned off.


\section{Space Weather Effects of Anthropogenic VLF Transmissions}
\label{sec:VLF}

\subsection{Brief History of VLF Transmitters}
\label{subsubsec:VLFhistory}

By the end of World War 1, the United States military began use of very low frequency radio transmissions (VLF; 3 - 30 kHz) for long-distance shore to surface ship communications \citep{Gebhard:1979a}.  Since very high power can be radiated from large shore-based antenna complexes, worldwide VLF communication coverage was feasible, and along with LF and HF systems (300 kHz - 30 MHz) these bands carried the major portion of naval communications traffic before later higher frequency systems came online.  Early experiments also showed that VLF could penetrate seawater to a limited depth, a fact realized by the British Royal Navy during World War I \citep{Wait:1977a}.  Given this realization, when the modern Polaris nuclear submarine era began in the 1950s, the US Naval Research Laboratory conducted a series of thorough radio propagation programs at VLF frequencies to refine underwater communications practices \citep{Gebhard:1979a}. Subsequent upgrades in transmission facilities led to the current operational US Navy VLF communications network, and other countries followed suit at various times.  For example, Soviet naval communication systems were likely brought online in the late 1920s and 1930s during the interwar expansion period, and high power VLF transmitters were later established in the late 1940s and 1950s for submarine communications and time signals.  These included Goliath, a rebuilt 1000 kW station first online in 1952 which partly used materials from a captured German 1940s era megawatt class VLF station operating at 16.55 kHz \citep{Klawitter:2000}.  

Table 2 of \citet{Clilverd:2009} lists a variety of active modern VLF transmitter stations at distributed locations with power levels ranging from 25 to 1000 kW.  These transmissions typically have narrow bandwidths ($<$ 50 Hz) and employ minimum shift keying \citep{Koons:1981}.  Along with these communications signals, a separate VLF navigation network (named Omega in the US and Alpha in the USSR) uses transmissions in the 10 kW range or higher \cite[e.g. Table 1 of][]{Inan:1984} with longer key-down modulation envelopes of up to 1 second duration.

\subsection{VLF Transmitters as Probing Signals}
\label{subsubsec:VLFtransmitters}

Beginning in the first half of the 20th century, a vigorous research field emerged to study the properties of VLF natural emissions such as whistlers, with attention paid as well to information these emissions could yield on ionospheric and magnetospheric dynamics.  Due to the high power and worldwide propagation of VLF transmissions, the geophysical research field was well poised to use these signals as convenient fixed frequency transmissions for monitoring of VLF propagation dynamics into the ionosphere and beyond into the magnetosphere [e.g. Chapter 2 of \cite{Helliwell:1965}; \cite[]{Carpenter:1966}].  This was especially true since VLF transmissions had controllable characteristics as opposed to unpredictable characteristics of natural lightning, another ubiquitous VLF source.  Beginning in the 1960s and continuing to the present, a vast amount of work was undertaken by the Stanford radio wave group and others (e.g. Yu. Alpert in the former USSR) on VLF wave properties, including transmitter reception using both ground-based and orbiting satellite receivers.  These latter experiments occurred both with high power communications and/or navigation signals and with lower power ($\sim$100 W), controllable, research grade transmitter signals.

The transmitter at Siple Station in Antarctica \citep{Helliwell:1988} is worthy of particular mention, as the installation lasted over a decade (1973--1988) and is arguably the largest and widest ranging active and anthropogenic origin VLF experiment series.  Two different VLF transmitter setups were employed at Siple covering 1 to $\sim$6 kHz frequency, with  reception occurring both in-situ on satellites and on the ground in the conjugate northern hemisphere within the province of Qu\'ebec.  Of particular note, the second Siple ``Jupiter'' transmitter, placed in service in 1979, had the unique property of having flexible high power modulation on two independent frequencies.  This allowed targeted investigations of VLF propagation, stimulated emissions, and energetic particle precipitation with a large experimental program employing a vast number of different signal characteristics not available from Navy transmitter operations.  These included varying transmission lengths, different modulation patterns (e.g. AM, SSB), polarization diversity, and unique beat frequency experiments employing two closely tuned VLF transmissions.  Furthermore, the ability to repeat these experiments at will, dependent on ambient conditions, allowed assembly of statistics on propagation and triggered effects.  These led to significant insights that were not possible for studies that relied on stimulation from natural waves (e.g. chorus) that are inherently quite variable.

Several excellent summaries of the literature on VLF transmission related subjects are available with extensive references, including the landmark work of \citet{Helliwell:1965} as well as the recent Stanford VLF group history by \citet{Carpenter:2015}.  As it is another effect of anthropogenic cause, we mention briefly here that a number of studies in the 1960s also examined impulsive large amplitude VLF wave events in the ionosphere and magnetosphere caused by above-ground nuclear explosions \cite[e.g.][]{Zmuda:1963,Helliwell:1965}.

Observations of VLF transmissions included as a subset those VLF signals that propagated through the Earth-ionosphere waveguide, sometimes continuing into the magnetosphere and beyond to the conjugate hemisphere along ducted paths \cite[]{helliwell:1958, Smith:1961}.  Ground based VLF observations \cite[]{Helliwell:1965} and in-situ satellite observations of trans-iono\-sphe\-ric and magnetospheric propagating VLF transmissions were extensively used as diagnostics. For example, VLF signals of human origin were observed and characterized in the topside ionosphere and magnetosphere for a variety of scientific and technical investigations with LOFTI-1 \cite[]{Leiphart:1962}, OGO-2 and OGO-4 \cite[]{Heyborne:1969, Scarabucci:1969}, ISIS 1, ISIS 2, and ISEE 1 \cite[]{Bell:1983}, Explorer VI and Imp 6 \cite[]{Inan:1977b}, DE-1 \cite[]{Inan:1982b, Inan:1984, Sonwalkar:1986, Rastani:1985}, DEMETER \cite[]{Molchanov:2006, Sauvaud:2008}, IMAGE \cite[]{Green:2005}, and COSMOS 1809 \cite[]{Sonwalkar:1994}.  VLF low Earth orbital reception of ground transmissions have been used also to produce worldwide VLF maps in order to gauge the strength of transionospheric signals \cite[]{Parrot:1990}. 

\subsection{VLF Transmitter Induced Effects in the Ionosphere and Magnetosphere}
\label{subsubsec:VLFinduced}

Given ambient terrestrial ionosphere and magnetosphere magnetic field and electron density values, VLF fixed frequency transmissions can not only pass through the medium but can under certain conditions trigger a variety of natural stimulated wave emissions, often interacting with existing particle populations in these regions.  Observations exploring these physical mechanisms have been and continue to be conducted both from the ground and in-situ within the ionosphere and magnetosphere. Section 3 of \cite{Parrot:1996} provides a mid 1990s snapshot review of observations and mechanisms.

Work in the 1960s and 1970s found that during sufficiently long key-down transmitter periods, triggered electron precipitation and stimulated wave events could occur, known as Trimpi effects [\cite[]{Helliwell:1973}; see also the observational and theoretical reviews by \cite{Helliwell:1988} and \cite{Omura:1991}]. These findings, other electron precipitation observations \citep{Inan:1984, Inan:1982a, Inan:1985, Imhof:1986, Vampola:1987, Vampola:1990, Sauvaud:2006} and later theoretical work \cite[e.g.][]{Abel:1998} led to the realization that VLF waves of sufficient amplitude might be useful in modifying the radiation belt electron populations through so-called Radiation Belt Remediation \citep{Inan:2003, Rodger:2006}, consisting of precipitation by alteration of particle pitch angle and therefore lowering of the minimum bounce altitude (2nd adiabatic invariant).  For example, the US Air Force DSX satellite scheduled for a 2017 launch will carry an onboard VLF transmitter for in-situ attempts to initiate radiation belt remediation mechanisms.

Unmodulated (i.e. carrier only) monochromatic waves from high power ground VLF transmitters have also been observed to quench natural magnetospheric hiss emissions \citep{Raghuram:1977}, and VLF transmitted pulses were thought to be amplified in the magnetosphere during storms through wave-electron interactions \citep{Smith:1991}. Ionospheric heating and perturbations in electron and ion densities have been associated with VLF transmitters as well \citep{Inan:1992,Parrot:2007}.

\subsection{VLF Transmitter Interactions Within the Radiation Belts}
\label{subsubsec:VLFobs}

The theoretical work of \citet{Abel:1998} predicted that the lifetimes of relativistic radiation belt electrons from L = 1.3 to 2.8 should be strongly influenced by VLF transmitter signals, and perhaps that this mechanism was responsible for the `slot' region between the inner and outer radiation belts, although updated VLF wave propagation modeling seemed to cast some doubt on these findings by suggesting a potential overestimation of wave amplitudes \citep{Starks:2008, Starks:2009, Cohen:2012}.  However, \citet{Abel:1998} and \citet{Kulkarni:2008} asserted that typical VLF transmitter frequencies could not resonate beyond L = 2.2 to 2.4 at the equator due to propagation characteristics and an increase in wave absorption as the wave frequency approaches the electron gyrofrequency. Significantly, these results were obtained using an electron density typical of conditions inside the plasmasphere ($10^3$ cm$^{-3}$).

The recent launch of the equatorial plane Van Allen Probes dual satellite mission in late 2012 \citep{Mauk:2013} has provided a large amount of new observational information on radiation belt electron populations and their dynamics.  The satellites carry a high fidelity electromagnetic wave package \citep{Kletzing:2013}, and data from this sensor has led to renewed study of the role of VLF transmission effects on relativistic electron dynamics within the magnetosphere.  It has been well established that electrons in the variable outer belt are affected by substorm injections and other inner magnetospheric dynamic processes, causing associated changes in populations at general ring current energies in the 10s to 100s of keV.  Through a multi-step process \cite[see Figure 1 of][]{Jaynes:2015}, recent research has found that these seed particles can provide significant energy reservoirs for highly efficient and dynamic wave-particle interactions on very short time scales, through mechanisms such as cyclotron resonance that directly affect relativistic electrons.  Resonant interactions can remove energy from pre-existing inner magnetospheric particle populations and can subsequently amplify coherent waves (e.g. rising tone chorus) that are highly efficient at affecting relativistic electrons through multiple mechanisms.  These factors result in large overall changes both in particle loss and/or acceleration for electrons from ring current energies out to ultra-relativistic energies.

New observations from Van Allen Probes wave and particle instruments have provided high fidelity equatorial plane in-situ measurements of ground based VLF transmissions out to $>$24,000 km radial distance (L$\sim$2.8), at the edge of what \citet{Foster:2016} term the VLF bubble.  VLF transmissions that propagate to these locations are strong and nearly omnipresent due to naval military operational considerations for submarine communications as mentioned previously.  Initial investigations point to these strong VLF fixed frequency transmissions of anthropogenic origin as playing a potentially important role in shaping dynamic responses of the radiation belts, primarily through interactions with wave amplification processes for those coherent plasma wave modes known to be highly resonant with relativistic particles outside the plasmasphere.  

In particular, the plasmapause and associated plasmasphere boundary layer \citep{Carpenter:2004} overlaps the region where the equatorial outward extent of the VLF bubble lies.  During disturbed intervals, the plasmapause can be retracted at or inward of the nominal VLF bubble edge, causing peak resonance energies at typical VLF transmission bands to shift dramatically away from the lower energies calculated by \citet{Abel:1998} and towards relativistic particle populations \citep{Foster:1976, Foster:2016}.  Significant spatial variation in resonance effects can therefore occur under retracted plasmasphere situations, with associated variations in wave-particle interactions possessing high degrees of coupling to relativistic electrons.  

\begin{figure}[h]
\begin{center}
\includegraphics[width=1\columnwidth]{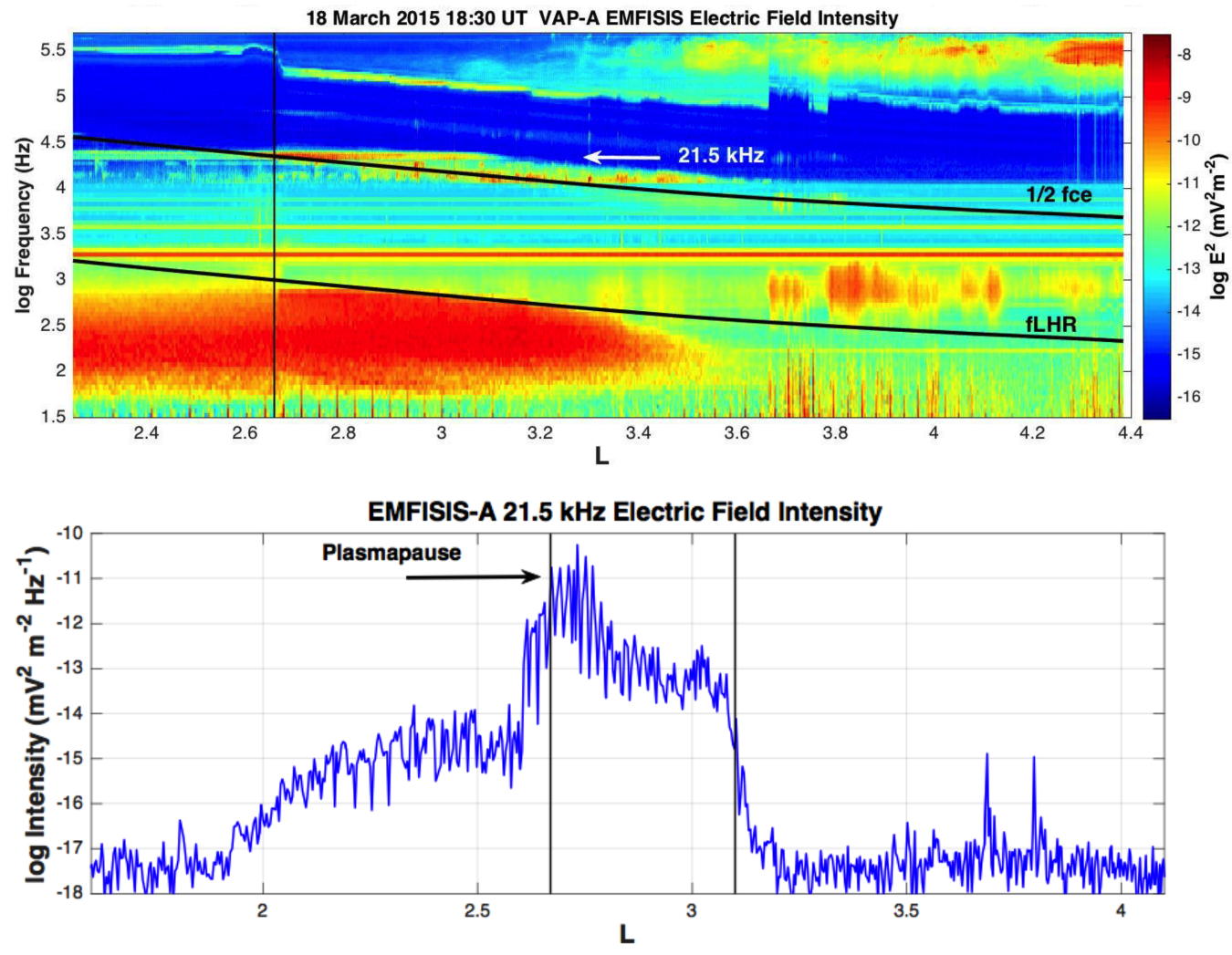}
\caption{
(Top) Van Allen Probes A EMFISIS electric field intensity spectra from 30 €‰Hz to 500 kHz are shown during a period following the 17 March 2015 great storm on 18 March 2015 at 18:30 UT, with 0.5 $f_{ce}$ and $f_{LHR}$ denoted by sloping black lines. The plasmapause boundary is indicated by the vertical black line at L = 2.65. In the region where a $\sim$21 kHz transmitter signal extended beyond the plasmapause, enhanced emissions were observed associated with frequencies close to a VLF transmitter band (white arrow).  (Bottom) Electric field wave intensity near $\sim$21 kHz increased by $> 10^5$ as L decreased from 3.2 to 2.7. The plasmapause position at L €‰= 2.65 determined from the EMFISIS upper hybrid frequency and the outer extent of enhanced emissions near the VLF transmitter frequency are indicated with black vertical lines. Adapted from \citet{Foster:2016}.
\label{fig:vlf_tx}
}
\end{center}
\end{figure}

\citet{Foster:2016} presented Van Allen Probes wave data from the great 17 March 2015 storm showing a dramatic increase in wave intensity around a $\sim$21 kHz VLF transmitter frequency band, in the region where the transmitter signal extended beyond the contracted plasmapause (cf. Figure~\ref{fig:vlf_tx}). These intensifications occurred at the radial location where the ground based VLF transmitter frequency bands were near half the local electron cyclotron frequency, under conditions appropriate for the generation of whistler-mode stimulated emissions. The resultant electric field intensity in the transmitter frequency band was observed up to 100,000 times background. Sharp gradients in highly-relativistic outer zone electron fluxes were also observed to be spatially coincident with the region of VLF enhancement, suggestive of the significant space weather effects of VLF transmitters in defining the earthward extent of outer radiation belt ``killer electrons'' \cite[e.g.][]{Baker:2014, Foster:2016}. In the region at the outer extent of the VLF bubble, beyond the contracted plasmapause, interactions involving signals from ground based VLF transmitters have a high cyclotron resonant potential to cause energetic electron precipitation, wave amplification, and/or other associated effects. 

We note in passing that these effects may potentially be congruent with previously reported observations of whistler mode amplification ($>$10X) in ground based VLF transmitter signals propagating through the magnetosphere along L = 2.5 field lines during large storms, with such observations interpreted by the authors as due to wave-particle interactions \citep{Smith:1991}. 

Investigations of these inner magnetospheric VLF transmitter stimulated effects near the VLF bubble edge are in early stages, and work is underway to examine the detailed relationship between VLF transmitter signals, stimulated emissions, and resonant particle populations.  Relevant for this discussion, wave-particle interactions in this category depend by their nature on the characteristics of pre-existing inner magnetospheric particle populations placed there by e.g. substorm injections or earlier disturbances with subsequent radial diffusion.  Further understanding of the potential anthropogenic space weather effects of ground based VLF transmitters therefore will depend on elucidation of the multi-step pathways involved as well as detailed examinations of the time history and characteristics of the required resonant particles. Nevertheless, VLF transmissions of anthropogenic origin may constitute a key space weather influence on pathways that fundamentally alter the storm-time radiation belt.  Under these assumptions, it is interesting for the reader to consider what the terrestrial radiation belt environment might have been in the pre-transmitter, and pre-observation, era.


\section{High Frequency Radiowave Heating}
\label{sec:HF}

Modification of the ionosphere using high power radio waves has been an important tool for understanding the complex physical processes associated with high-power wave interactions with plasmas. There are a number of ionospheric heating facilities around the world today that operate in the frequency range $\sim$2--12 MHz. The most prominent is the High Frequency Active Auroral Research Program (HAARP) facility in  Gakona, Alaska. HAARP is the most powerful radio wave heater in the world; it consists of 180 cross dipole antennas  with a total radiated power of up to 3.6 MW and a maximum effective radiated power (EFR) of  $\sim$4 GW. The other major heating facilities are EISCAT, SURA, and Arecibo. EISCAT is near Tromso, Norway and has an EFR of $\sim$1 GW. SURA is near Nizhniy Novgorod, Russia and is capable of transmitting $\sim$190 MW ERP. A new heater has recently been completed at Arecibo, Puerto Rico with  $\sim$100 MW ERP. There was a heating facility at Arecibo that was operational in the 1980s and 1990s but it was destroyed by a hurricane in 1999.  The science investigations carried out at heating facilities span a broad range of plasma physics topics involving ionospheric heating, nonlinear wave generation, ducted wave propagation, and ELF/VLF wave generation to name a few. 

During experiments using the original Arecibo heating facility,  \cite{Bernhardt:1988} observed a dynamic interaction between the heater wave and the heated plasma in the 630nm airglow: the location of HF heating region changed as a function of time.  The heated region drifted eastward or westward, depending on the direction of the zonal neutral wind,  but eventually ``snapped back'' to the original heating location. This was independently validated using the Arecibo incoherent scatter radar for plasma drift  measurements \cite[]{Bernhardt:1989}. They suggested that when the density depletion was significantly transported in longitude, the density gradients would no longer refract the heater ray and the ray would snap back, thereby resulting in a snapback of the heating location as well. However, a recent simulation study using a self-consistent first principles ionosphere model found that the heater ray did not snap back but rather the heating location snapped back because of the evolution of the heated density  cavity \cite[]{Zawdie:2015}.

The subject of ELF wave generation is relevant to communications with submarines because these waves penetrate sea water. It has been suggested that these waves can be produced by  modulating the ionospheric current system via radio wave  heating \cite[]{papadopoulos:1989}. Experiments carried out at HAARP \cite[]{Moore:2007} demonstrated this by sinusoidal modulation of the auroral electrojet under nighttime conditions. ELF waves were detected in the Earth's ionosphere waveguide over 4000 km away from the HAARP facility.

VLF whistler wave generation and propagation have also been  studied with the HAARP facility. This is important because whistler waves can interact with high-energy radiation belt electrons. Specifically, they can pitch-angle scatter energetic electrons into the loss cone and precipitate them into the  ionosphere \cite[]{Inan:2003}. One interesting finding is that the whistler waves generated in the ionosphere by the heater can be amplified by specifying the frequency-time format of the heater, as opposed to using a constant frequency \cite[]{Streltsov:2010}. 

New observations were made at HAARP when it began operating at its maximum radiated power 3.6 MW. Specifically, impact ionization of the neutral atmosphere by heater-generated suprathermal electrons can generate artificial aurora observable to the naked eye \cite[]{Pedersen:2005} and a long-lasting, secondary ionization layer below the $F$ peak \cite[]{Pedersen:2009}. The artificial aurora is reported to have a ``bulls-eye'' pattern which is a refraction effect and is consistent with ionization inside the heater beam. This phenomenon was never observed at other heating facilities with lower power (e.g., EISCAT, SURA).


\section{Chemical Releases}
\label{sec:release}

Chemical releases in space have been carried out since the early 1960s. The first release experiments were performed by scientists at the Max Planck Institute for Extraterrestrial Physics. Barium was used because it rapidly ionizes in sunlight. One of the first successful experiments was at an altitude of 2000 km \cite[]{Luest:2001} but most of the subsequent experiments were in the altitude range 150 -- 250 km. The original purpose of these experiments was to create an artificial cometary tail in order to understand the interaction of the solar wind with cometary ions \cite{Biermann:1951},  however, it was soon realized that these releases could be used as a diagnostic of the ionosphere (e.g., the neutral wind and electric field), as well as a method to study basic plasma physics processes (e.g., ion-neutral interactions, plasma wave generation). Recognizing the scientific value of chemical release experiments, two chemical release missions were carried out by NASA: AMPTE (Active Magnetospheric Particle Tracer Explorers) \cite[]{Haerendel:1985}  and CRRES (Combined Release and  Radiation Effects Satellite) \cite[]{Reasoner:1992}. 

AMPTE was launched in August, 1984. Lithium and barium were released in  the solar wind and in the geomagnetic tail. A key objective of the solar wind releases echoed the original motivation of chemical releases: the generation and evolution of an artificial co\-me\-tary tail. An interesting and surprising result was that the barium ion cloud did not immediately move in the direction of the solar wind, rather, the cloud initially moved perpendicular to the IMF and solar wind direction. Subsequent analysis indicated that barium ion inertial length was sufficiently large that the Hall term dominated the  early evolution of the cloud \cite[]{Harold:1994}. A second objective was to trace lithium ions from the solar wind into the magnetosphere. However, analysis of data from the Charge Composition Explorer (CCE) inside the magnetosphere following two lithium releases detected no lithium in the magnetosphere \cite[]{Krimigis:1986}. AMPTE barium releases in the earth's magnetotail revealed complex motions \cite[]{Mende:1989} and rapid plasma  structuring \cite[]{Bernhardt:1987}. The structuring was explained by a new Rayleigh-Taylor instability mediated by the Hall term \cite[]{Hassam:1987}.

The CRRES satellite was launched in July, 1990 and the chemical releases were from the CRRES satellite itself as well as from a number of sounding  rockets. The original mission plan was to launch the satellite from the space shuttle. It would spend 90 days in low Earth orbit (LEO) and perform chemical release experiments. It would then  transfer to a geosynchronous orbit and perform more releases.  However, because of the 1986 Challenger disaster the mission plan changed and the satellite was inserted into a geosynchronous transfer orbit (GTO). To accommodate the loss of the LEO chemical releases, sounding rockets were used. A number of diverse science objectives were addressed by the mission: critical ion velocity ionization,  generation and evolution of diamagnetic cavities, magnetosphere-ionosphere coupling, modification of energetic electron distribution functions, stimulation of magnetospheric waves, and possible inducement of enhanced  auroral activity \cite[]{Reasoner:1992}. The bulk of the chemical releases used lithium and barium. However, one rocket experiment used sulfur hexafluoride (SF$_6$). SF$_6$ reacts with cold electrons and becomes a negative ion. The idea was to release SF$_6$ in the post-sunset, bottomside, equatorial ionosphere to create a large electron `hole' which would act as a trigger for equatorial spread $F$. Equatorial spread $F$ did occur during this experiment and believed to be triggered by the chemical release (Mendillo, private communication, 2016) but the cause was not unambiguous.

Another chemical that has been used extensively to detect and measure thermospheric winds in the altitude range 80 -- 140 km is trimethyl aluminum (TMA). TMA reacts with oxygen to produce chemiluminescence that is easily observable at night. A large database of wind profiles has been developed \cite[]{Larsen:2002} that show variability in thermospheric winds. as well  as strong wind  shears in the E-region between 100 and 110 km that are often  associated with sporadic E.

Recently the United States Air Force conducted chemical release experiments at the Kwajalein Missile Range to assess their impact on radio wave propagation in the ionosphere. The Metal Oxide Space Cloud (MOSC) experiment was carried out in 2013 in which samarium vapor was released at 170 km and 180 km. The samarium vapor created an ionized plasma cloud through both photoionization and chemical ionization; this cloud created observable signatures in optical sensors, radar backscatter, and HF propagation diagnostics. A major goal of this experiment is to determine the viability of using chemical releases to maintain communications during disturbed ionospheric conditions by suppressing ionospheric instabilities. In fact the Air Force is sponsoring research to develop small satellites known as CubeSats to release chemicals to improve radio frequency communications \cite[]{Hambling:2016}.

In summary, chemical releases have been used for over 50 years to study plasma processes in the space environment and still continue to be used. The two overarching objectives of these experiments  are to diagnose the environment (e.g., measure physical quantities such as thermospheric winds and electric fields) and to artificially generate plasma processes to better understand them (e.g., plasma waves, diamagnetic cavities, ion-neutral interactions). 


\section{Space Debris}
\label{sec:debris}

Space debris consists of human-made waste, junk, and defunct objects orbiting in space \cite[]{Klinkrad:2010}.   The debris consists of objects as large as spent rocket stages, and as small as flakes of paint.  Additional debris can be produced by collisions between debris objects.    Some debris has been produced by intentional tests of anti-satellite systems.  As the use of space has increased dramatically since Sputnik, the amount of human space debris has also increased substantially.   The space debris literature often discusses whether the amount of debris, and its interactions with itself, has reached the ``Kessler syndrome'' \cite[]{Kessler:1978}, where the density of debris in some orbital locations becomes high enough that debris collisions will cascade.   If so, some orbits could become unusable.

Space debris can be considered an anthropogenic space hazard.  But debris is not a space weather phenomenon.  Nevertheless, space debris can damage and even destroy space assets, just as can naturally-occurring micrometeoroids.   Debris can ``sand-blast'' and pock-mark surfaces in space, from thermal blankets to solar cells and optical telescopes and star trackers.  Large debris objects can potentially render an entire mission inoperable through a collision.   Debris in the lowest orbits will be de-orbited and burned up in the atmosphere during times of solar maximum when the atmosphere increases at these altitudes, causing increased drag on the debris.  The removal of debris from higher orbits, including at geosynchronous where many communications and national security space assets fly, remains an important engineering challenge.   


\section{Summary Discussion}
\label{sec:summary}

While only recognized in recent times, human intervention in the Earth's space environment has occurred ever since the decision was made for alternating current as the electrical grid power source.    These interventions, both inadvertent (such as the emission of power line radiation into the magnetosphere) and intentional (such as high altitude nuclear explosions) can produce a range of effects that not only resemble, but in many cases are significantly more severe, than what can be expected from naturally occurring space weather events.  As described in this paper, some anthropogenic effects have been demonstrated to have the potential for globally destructive results.

The sections in this paper that describe the phenomenology of high altitude nuclear explosions, the actual formation and phenomenology of long-lasting artificial radiation belts, their damaging effects on contemporary satellites and their observable geophysical effects present a picture from the past, but with a magnified contemporary relevance in the context of space weather.  These early, unrepeatable experiments remain an important reference point in understanding the phenomena that a disturbed space environment can present -- nowadays expected to be the result of solar activity, rather than human intervention.

It is remarkable how the early space pioneers, within the first decade of the space age, were able to make progress in understanding the Earth's space environment, thanks to increasingly sophisticated and targeted instrumentation on a range of Earth satellites, as well as taking advantage of the man-made space weather events.  The situation currently is well exemplified by the successes of the detailed observations of the dual Van Allen Probes in the radiation belts, as well as by the measurements of numerous other satellite missions all contributing to building an ever more detailed picture of phenomena in the near-Earth space environment.  At the same time, modeling and simulations have progressed to an incomparable extent since the early means at the disposal of the observers and theoreticians of the Starfish events in 1962.

Current human intervention that can affect the space environment is rather more benign, such as using VLF signals as probes of near-Earth phenomena and ionospheric heating for clarifying the complex interactions of the ionosphere with many different wave fields.    These experiments are carried out with a reasonable understanding of their potential effects and present only well-contained risks.  Other experiments, such as the chemical releases that have been performed in the ionosphere and magnetosphere, have -- generally  speaking -- negligible and evanescent effect on the environment, while contributing in a more limited way, to understanding relatively small-scale dynamic processes.

James Van Allen could rightly claim that the Argus nuclear explosions in 1958 ``... undoubtedly constitute the greatest geophysical experiment ever conducted by man. The observations have great significance in understanding the nature of the geomagnetic field, the mechanisms of trapping of particles in the geomagnetic field, the origin of aurorae and-magnetic storms, and the density of the very high atmosphere.'' (Letter from James Van Allen to James R. Killian, Office of the President of the United States, dated February 21, 1959.)  It is to be hoped that no high altitude nuclear explosions will be carried out in the future, as the reliance of civilization in this first part of the 21st century on a complex technology infrastructure in the near-Earth space environment has increased enormously when compared to the situation half a century ago.  In fact the ``experiments'' in 1958, originally designed for mitigation purposes, together with the rather more destructive ones in 1962, did bring a better early understanding of the Earth's natural radiation belts.  Since that era, near-Earth space research has progressed to a far greater level of sophistication so that such experiments are not only unthinkable, but they would destroy much of the infrastructure of our current civilization.

The most dramatic -- and unexpected -- anthropogenic space weather phenomenon is the high-altitude nuclear EMP (HEMP). As it can be seen in \figurename~\ref{fig:EMPphases} a megaton-class nuclear explosion at an altitude of $\sim400$ km produces a very fast electromagnetic pulse. The rise-time of this pulse is about 100 times faster than the rise-time of lightning while the generated electromagnetic fields are comparable ($\sim50$ kV/m). In addition, this pulse covers a continent-size area at the same time. The effect can be devastating for modern society: it can disrupt transportation, energy production and distribution, global and local computer networks and many other aspects of modern life we take for granted. The Hiroshima and Nagasaki nuclear devices had explosive powers of 15 kt and 21 kt, respectively. A relatively crude, but factor of 20 larger device can cause devastating effects in the US or Western Europe. Such a device in the hands of terrorists or a rogue nation would pose a major threat to the world.

While the study of space weather, as a scientific discipline, has deep roots from the time of installation of the first telegraph lines, to the early years of the space age, the subject remains a lively and timely topic that has become essential as part of the global technological infrastructure in space and on the ground.  Anthropogenic intervention forms a portion of the past, and a modest component of the present, but it is now rightly dominated by the natural phenomena from the Sun to the Earth.


\begin{acknowledgements}
The authors thank the International Space Science Institute, Bern, Switzerland and its staff for organizing and supporting the Workshop on the Scientific Foundations of Space Weather that motivated the work in this paper. The work performed at the University of Michigan was supported by National Science Foundation grant AGS-1322543. JDH was supported by NRL Base Funds. Work at the Massachusetts Institute of Technology was sponsored by US National Science Foundation grant AGS-1242204. Work at the University of Colorado/LASP was supported by funding from NASA and the National Science Foundation. The authors thank Vaughn Hoxie, Scot Elkington, Hong Zhao, and Tom Mason for extraordinary efforts in adapting and portraying data from the Explorer XV and Van Allen Probes missions.
\end{acknowledgements}

\bibliographystyle{aps-nameyear}     
\bibliography{anthro-sw}

\end{document}